\documentstyle[epsfig]{mn}
\title[The UV Continuum Excess in Radio Galaxies]
{The origin of the UV excess in powerful radio galaxies: spectroscopy and polarimetry 
of a complete sample of intermediate redshift radio galaxies}
\author[Tadhunter et al.]
{C. Tadhunter$^1$, R. Dickson$^{2}$, R. Morganti$^3$, T.G. Robinson$^1$, K. Wills$^1$,  
\newauthor M. Villar-Martin$^4$, 
M. Hughes$^4$
\\
$^{1}$Department of Physics and Astronomy, University of Sheffield, 
Hounsfield Road, Sheffield, S3 7RH, UK\\ 
$^{2}$Jodrell Bank Observatory, University of Manchester, Macclesfield, 
Cheshire, SK11 9DL. \\
$^{3}$Netherlands Foundation for Research in Astronomy, Postbus 2, 7990 AA Dwingeloo, 
The Netherlands \\
$^{4}$ Division of Physical Sciences, University of Hertfordshire, Herts}

\date{}
\def\ltsim{\ifmmode\stackrel{<}{_{\sim}}\else$\stackrel{<}{_{\sim}}$\fi}
\def\gtsim{\ifmmode\stackrel{>}{_{\sim}}\else$\stackrel{>}{_{\sim}}$\fi}
\begin{document}
\maketitle
\begin{abstract}
We present spectroscopic and polarimetric observations of a
complete, optically unbiased sample of 2Jy radio galaxies at intermediate redshifts ($0.15 
< z < 0.7$). These data --- 
which cover the nuclear regions of the target galaxies --- allow us to 
quantify for the first time the various components that contribute to the UV excess in the 
population of powerful, intermediate redshift radio galaxies. We find that, 
contrary to the results of previous 
surveys --- which have tended to be biased towards the most luminous and spectacular 
objects in any redshift range --- the contribution of scattered quasar light to the UV excess 
is relatively minor in most of the objects in our sample. Only 7 objects (32\% of the complete
sample) show significant polarization in the rest-frame UV, and none of the 
objects in our sample is polarized in the near-UV at the $P > 10$\% level. Careful 
measurement and modelling of our spectra have allowed us to quantify the contributions of 
other components to the UV excess. We show that nebular continuum (present in all 
objects at the 3 --- 40\% level), direct AGN light (significant in 40\% of objects), and 
young stellar populations (significant in 15 --- 50\% of objects) 
all make  important contributions to the 
UV continuum in the population of powerful radio galaxies. These results serve to emphasise 
the multi-component nature of the UV continuum in radio galaxies. The
results also point to an interesting link
betweeen the optical/UV and far-IR properties of our sample objects, in the
sense that the objects with the clearest evidence for optical/UV starburst activity
are also the most luminous at far-IR wavelengths. This supports the idea that
the cooler dust components in radio galaxies are heated by starbursts rather than
by AGN. 


\end{abstract}

\section{Introduction}

Given the large look back times encompassed by the most distant radio sources, one 
motivation for studying such objects is their potential use as probes of the formation and 
evolution of giant early-type galaxies in the early universe. However, all studies aimed at 
using radio galaxies in this way have to face the problem of distinguishing the effects of 
the AGN and radio jet activity from genuine signs of galaxy evolution. 
This problem is particularly acute in the case of studies of the 
continuum properties. Compared with normal early-type galaxies, powerful radio 
galaxies can show continuum excesses at both optical/UV (e.g. Lilly \& Longair
1984, Smith \& Heckman 1989) and far-IR/sub-mm wavelengths
(Golombek et al. 1988, Heckman et al. 1994, Archibald et al. 2001). Therefore, a key 
issue for these objects is whether these continuum excesses are a consequence of 
recent star formation which may be linked to evolutionary processes in the early-type 
host galaxies or, given that these objects contain powerful AGN and radio jets, a direct 
consequence of the activity.

The presence of a UV excess in the continua of radio galaxies was first demonstrated by 
the photometric observations of samples of high redshift ($z > 0.5$) radio galaxies in the 
early 1980's. These observations showed evidence for bluer optical-IR colours than 
expected for non-evolving or passively evolving elliptical galaxies (e.g. Lilly \& Longair 
1984). 
Initially, the UV excess was interpreted in terms of bursts of star formation, possibly 
linked to the evolution of the host galaxies. This interpretation is attractive in the light of 
morphological studies which show evidence for recent mergers in a large fraction of 
powerful radio galaxies at low redshifts (Heckman et al. 1986); and merger-induced 
star formation has been 
suggested as a possible triggering mechanism (Smith \& Heckman 1989). 
However, given the degree of nuclear and extranuclear activity likely to be present in most
powerful radio galaxies, some caution is required in deducing starburst
properties purely on the basis of broad-band photometric measurements. 
 
Recognising the potential AGN contribution, an 
alternative explanation for the UV excess 
was stimulated by the development of the anisotropy-based 
unified schemes in the late 1980's (e.g. Barthel 1989). In the frame of such schemes the UV excesses
can be explained in terms of light scattered from broad radiation cones 
of the hidden quasar nuclei (Tadhunter et al. 1988, Fabian 1989). 
Early polarimetric attempts to test this 
model proved successful in the sense that they showed the high degrees of linear 
polarization characteristic of anisotropic scattering in the UV continua of several high 
redshift radio galaxies (e.g. Tadhunter et al. 1992, Cimatti et al. 1993, 
Vernet et al. 2001). However, while these observations demonstrate that scattered 
quasar light is an important component of the UV continuum in {\it some}  sources, they 
do not establish the significance of the scattered component in the general population of 
powerful radio galaxies. Because polarimetric observations of faint 
objects are difficult, previous studies have tended to be biased towards the 
brightest, most spectacular objects in a given redshift range. There are also 
redshift-dependent biases which arise because optical (mostly V-band) observations 
sample the rest-frame UV in the high redshift objects --- with minimal dilution by the old 
stellar populations of the host galaxies --- but sample the rest-frame optical in the low 
redshift objects --- with substantial dilution by the old stellar populations. The importance 
of this observational selection effect is emphasised by multi-wavelength polarimetric 
observations of individual sources which show a sharp decline in the measured 
polarization between the UV and the optical (Tadhunter et al. 1996, Ogle et al. 1997, 
Tran et al. 1998). 

In addition to the scattered component, detailed observations over the last decade have 
revealed the presence of two further activity-related components which can contribute to 
the UV excess. These are: the nebular continuum emitted by the extended emission line 
nebulae (Dickson et al. 1995); and direct AGN light emitted by weak, or partially extinguished, 
quasars in the 
nuclei of the galaxies (Shaw et al. 1995). The nebular continuum is likely to be particularly 
significant in 
regions where the emission lines have large equivalent widths, including the extended 
emission line nebulae around powerful radio galaxies. In contrast, the direct AGN 
component will only be important in the nuclear regions of the sources. 

Most recently, events have turned full circle with the spectroscopic detection of young stellar 
populations in at least some powerful radio galaxies (e.g. Tadhunter et al. 1996, Melnick 
et al. 1997). The detection of this component is consistent with the early interpretation of the 
UV excess in terms of starbursts associated with the evolution of the host galaxies (Lilly 
\& Longair 1984). Unfortunately, apart from cases in which it dominates the optical 
continuum (e.g. Miller 1981), the starburst component is notoriously difficult to detect at optical 
wavelengths. Its presence can be masked by the light of the old stellar 
populations in the bulges of the host galaxies, by the various activity-related continuum 
components noted above, and by emission lines which can contaminate the absorption 
features characteristic of young stars. This is illustrated by the case of 3C321 which 
shows polarimetric evidence for a significant scattered quasar component, but also shows 
evidence for a starburst component in the form of a Balmer break and Balmer absorption 
features (Tadhunter et al. 1996, Robinson et al. 2000). It is notable that the starburst 
component in 3C321 only came to light through detailed modelling of the optical/UV 
continuum using a combination of spectrophotometry and spectropolarimetry 
measurements.

Given the complex circum-nuclear environments of powerful radio 
galaxies revealed by recent HST imaging studies (e.g. Jackson, Tadhunter \& Sparks 
1998), it is not surprising that no single mechanism is responsible
for the UV excess. Observations of individual sources demonstrate the presence of at least four 
UV-emitting components that can contribute to the UV excess: scattered AGN light, direct 
AGN light, nebular continuum, and the light of young stellar populations. However, the relative 
importance of these components, and particularly
the importance of any starburst component, is not clear from the previously published data. In 
this paper we attempt to remedy this situation by combining spectroscopic and 
polarimetric observations to quantify the contributions of the various UV-emitting 
components in a complete, optically unbiased sample of powerful 2Jy radio galaxies at
intermediate redshifts ($0.15 < z < 0.7$). We also consider the link
between the optical/UV signs of star formation activity and the far-IR continuum
excess. In a companion paper we report a similar study
of a lower redshift sample of 3C radio galaxies ($z < 0.2$: Wills et al. 2002).

Throughout this paper we assume a Hubble constant of $H_0 = 50$ km s$^{-1}$ Mpc$^{-1}$ and
a deceleration parameter of $q_0 = 0$.
    
\section{Sample selection}

The objects included in this study comprise radio galaxies selected from the Tadhunter et al.(1993)  complete 
sample of 2Jy radio sources with redshifts $z < 0.7$ and declinations $\delta < +10$, which is 
itself a subsample of the Wall and Peacock (1985) sample of radio sources with flux 
densities greater than 2Jy at 2.7 GHz. As discussed in Tadhunter et al. (1993, 1998) the 
$z < 0.7$ sample has a high level of completeness. Low S/N optical spectra and 
identifications for all the $z < 0.7$ sample are presented in Tadhunter et al. (1993) and di 
Serego Alighieri  et al. (1994); radio maps for this sample are presented in Morganti et al. 
(1993, 1999); and X-ray observations are presented in  Siebert et al. (1996). 
Discussion of the radio observations in the context of the unified schemes can be found in 
Morganti et al. (1995, 1997), while a discussion of the nature of the correlations between radio 
and optical emission line properties of the $z < 0.7$ sample is presented in Tadhunter et al. 
(1998).

\begin{table} 
\begin{center}
\renewcommand{\thefootnote}{thempfootnote} 
\begin{tabular}{llcll} \hline \hline 
{\bf Object} & {\bf Other name} & {\bf z} & {\bf Radio } & {\bf Optical }    \\ 
 & & & {\bf Morph.} &{\bf Spect.} \\\hline
0023--26     &  OB-238          &  0.322  & CSS & NLRG          \\
0035--02     &  3C17            &  0.220  & II  & BLRG        \\
0038+09     &    3C18          &  0.188  & II  & BLRG         \\
0039--44     &                  &  0.346  & II  & NLRG         \\
0105--16     &  3C32            &  0.400  & II  & NLRG         \\
0117--15     &  3C38            &  0.565  & II  & NLRG         \\
0235--19     &  OD-159          &  0.620  & II  & (BLRG)        \\
0252--71     &                  &  0.566  & CSS & NLRG         \\
0347--05     &  4C05.16         &  0.339  & II  & BLRG           \\
0409--75     &                  &  0.693  & II   & NLRG        \\
1306--09     &  OP-10           &  0.464  & CSS  & NLRG        \\
1547--79     &                  &  0.483  & II   & BLRG       \\
1549--79     &                  &  0.150  & FSC  & NLRG       \\
1602+01      &  3C327.1         &  0.462  & II   & BLRG       \\
1648+05      &  Her A           &  0.154  & I/II & WLRG       \\
1932--46     &                  &  0.231  & II   & BLRG       \\
1934--63     &                  &  0.183  & GPS  & NLRG       \\
1938--15     &                  &  0.452  & II   & BLRG       \\
2135--20     &  OX-158          &  0.635  & CSS  & (BLRG)       \\
2211--17      &  3C444           &  0.153  & II   & WLRG       \\
2250--41     &                  &  0.310  & II   & NLRG       \\
2314+03      &  3C459           &  0.220  & II   & NLRG       
\end{tabular}
\end{center}
\caption[Details of the  objects studied.]
{Details of the sources in the complete sample of radio sources discussed in this
paper. Column 4 gives the radio morphologies of the sample objects, based on the maps 
from Morganti et al. (1993), and Morganti et al. (1999) (key: II -- Fanaroff Riley class II;
I -- Fanaroff Riley class I; CSS -- compact steep spectrum ($D < 15kpc$); FSC -- flat spectrum
core source; GPS --- gigahertz peak spectrum radio source).  The optical spectral
classifications in column 5 are based on the deep spectra discussed in  this paper and Tadhunter et al. (1998) 
(key: NLRG -- narrow line radio galaxy; BLRG -- broad line
radio galaxy; WLRG -- weak line radio galaxy). Objects for which the classification is  uncertain
are indicated by brackets.}    
\label{table:objects}
\end{table}

Although in section 4 below we will consider the UV excess in the $z < 0.7$ sample of Tadhunter
et al. 1993 as a whole, most of this paper will be concerned with observations of a more
restricted sample, selected to facilitate a more detailed 
study of the optical/UV continuum properties. The additional selection criteria for this
restricted sample are: redshifts in the range $0.15 < z < 0.7$ and
right ascensions in the range $13^{hr} < RA <  05^{hr}$. The lower redshift limit for this restricted
sample was 
chosen to ensure that the B-band filter used for the polarimetry observations covers the 
rest-frame UV for all the sample objects, while the RA restriction was intended to ensure 
that we observed a complete RA-limited sample of managable size, given the constraints 
on the observing time. Some properties of the resulting sample of 22 objects are 
presented in Table 1. 

\begin{table} 
\begin{tabular}{l|llccl}  \hline \hline
{\bf Object} & {\bf Date} &
  {\bf S/W} & {\bf PA ($^{\circ}$)} & {\bf Time (s)} \\ \hline
0023---26 & 22/7/93 & 1.7/2 & 270 & 900(R + B)  \\ 
0035---02 & 21/7/93 & 1.4/2 & 270 & 900(R)  600(B)  \\
0038+09  & 22/10/95 & 1.7/5 & 270 & 900(B)  \\  
0039---44 & 23/7/93 & 1.5/2 & 270 & 900(R) 1200(B) \\  
0105---16 & 23/7/93 & 1.5/2 & 270 & 900(B + R) \\ 
0117---15 & 23/7/93 & 1.5/2 & 270 & 900(B + R) \\ 
0235---19 & 20/10/95 & 1.5/5 & 270 & 2$\times$900(B) \\ 
         & 20/10/95 & 1.5/5 & 270 & 600(B) \\ 
0252---71 & 21/7/94  & 1.5/5 & 270 & 600(B)  \\ 
         & 21/7/94  & 1.5/5 & 270 & 900(R) \\ 
0347+05  & 22/10/95 & 1.5/5 & 270 & 600(B) \\ 
         & 22/10/95 & 1.5/5 & 270 & 300(R) \\ 
0409---75 & 22/10/95 & 1.4/2 & 208 & 700(B) \\ 
         & 22/10/95 & 1.4/2 & 208 & 600(R) \\ 
1306---09 & 23/7/93  & 1.3/2 & 270 & 900(R + B) \\ 
         & 12/7/94  & 1.8/2 & 270 & 900(R + B) \\ 
1547---79 & 20/7/93  & 1.2/2 & 270 & 900(R + B) \\
         & 22/7/93  & 1.3/2 & 270 & 900(B) \\
         & 23/7/93  & 1.2/2 & 270 & 900(B) \\ 
1549---79 & 12/7/94  & 1.9/5 & 270 & 1200(B) \\ 
1602+01  & 30/7/92  & 1.5/1.6 & 32 & 2$\times$1200(R) \\ 
         & 30/7/92  & 1.5/2 & 32 & 1200(R) \\ 
1648+05  &  12/07/94 & 1.6/2 & 300 & 900(R + B) \\ 
1932---46 & 22/7/93  & 1.7/2 & 270 & 900(R + B) \\
         & 23/7/93  & 1.5/2 & 270 & 600(R + B) \\ 
1934---63 & 11/7/94  & 2.2/2 & 270 & 1200(B) \\ 
1938---15 & 22/7/93  & 1.3/2 & 211 &  900(R + B) \\  
2135---20 & 23/7/93  & 1.5/2 & 270 & 1200(B) \\ 
         & 30/7/92  & 1.9/1.6 & 343 & 1200(R) \\ 
         & 30/7/92  & 1.9/2.0 & 343 & 3$\times$1200(R + B) \\
2211---17  & 21/10/95 & 1.1/2   & 225 & 900(B)  \\ 
2250---41 & 20/7/93  & 1.9/2 & 270 & 2$\times$1200(B) \\
         & 20/7/93  & 1.9/2 & 270 & 900(R)  \\
         & 22/7/93  & 1.6/2 & 270 & 1200(B) \\ 
2314+03  & 22/7/93  & 1.6/2 & 270 & 1200(B) \\ 
         & 22/7/93  & 1.6/2 & 270 & 600(R) \\ 
\end{tabular}
\caption[Summary of  the spectroscopy of the  complete sample.]
{Observational details of the spectroscopic observations for the objects in Table 1.
Column 3 lists the seeing FWHM (S) and slit width (W) in arcseconds  for the observations. 
Column 4  lists the position angle of the slit for each observation. 
Column 5 gives the integration time in seconds for each observation. All of the southern sample of radio galaxies
were observed  using EFOSC with  the R300 and B300 gratings, with the exception of 1602+01 (3C327.1)
which was observed with the red arm of the ISIS spectrograph on the 4.2m WHT telescope (30/7/92), 
and 2135-20, which was observed with the B300 grating of EFOSC (23/7/93 run) and with both
the red and blue arms of ISIS on the WHT telescope. }
\label{table:obs-spec-samp}
\end{table}

The overall aim of our survey was to obtain accurate B-band polarimetry measurements 
and deep spectra covering at least the rest wavelength range 3500 --- 5500\AA \, for all of the 
restricted sample in Table 1. In reality, while we succeeded in obtaining deep spectra for 
the whole sample, time restrictions meant that we failed to obtain B-band polarimetry 
observations for three (15\%) of the objects: the CSS source PKS0252-17, and the weak 
lined radio galaxies PKS1648+05 (Her A) and PKS2211-17(3C444). In the case of  PKS0252-71
contamination by the light of a foreground galaxy precludes detailed study of the UV 
continuum properties, while for PKS1648+05 and PKS2211-17 it was felt that, considering 
of the lack of emission line activity, it was unlikely that any UV excess could be due to a 
polarized (scattered AGN) component. Failure to observe these three objects does not 
significantly affect our conclusions. 

The key advantages of this survey for studying the UV excess are: 
(a) we have both polarimetric and spectroscopic observations for the overwhelming majority 
of the objects in Table 1, so our survey is not biased towards the brightest, 
most spectacular objects in the chosen redshift range; and 
(b) we have reduced the problem of varying dilution of any scattered component by the 
old stellar populations by observing all the objects in the rest-frame UV. 

Some of the polarimetric and spectroscopic results from the survey have already been 
published in Shaw etal. (1995) and Tadhunter et al. (1994). Here we present results for 
the remainder of the sample in Table 1, and discuss the results for this sample 
collectively. 

\section{Observations and reductions}

B-band polarimetry data and long-slit spectra for most of the objects in Table 1 were 
obtained in three runs on the ESO 3.6m telescope at La Silla in Chile between 1993 and 
1995. All the data from these runs were taken with the ESO Faint Object Spectrograph and 
Camera No.1 (EFOSC1) with a thinned TEK chip (No. 26), resulting in an angular scale 
of 0.62 arcseconds/pixel.

Long-slit spectra for the sample were obtained using the R300 and B300 grisms in 
EFOSC1 (see Melnick et al. 1989). These grisms together provide complete coverage of 
the observed wavelength range 3600 to 9900\AA\, with a dispersion of 6.2\AA/pixel for the 
B300 grism and 7.3\AA/pixel for the R300 grism. The instrumental resolution was $\sim$20\AA\, 
with the 2 arcsecond slit. With this coverage it was possible to cover at least the rest 
wavelength range 3500-5500\AA\, for all of the objects observed with EFOSC1. Details of the
spectroscopic observations are given in Table 2.

In addition to the spectra obtained with EFOSC for most of the sources, two of the 
objects --- 3C327.1 and PKS2135-20 --- were observed using the 4.2m William Herschel 
Telescope (WHT) on La Palma with the ISIS dual-beam spectrograph. Both objects were observed 
using the R158R grating on the red arm of ISIS with an EEV CCD detector and a GG495 
order sorting filter, yielding a dispersion of 2.72\AA/pixel and a resolution of 10\AA\, for a 1.6 
arcsecond slit. 


\begin{table} 
\begin{center}
\begin{tabular}{l|lccc} \hline \hline
{\bf Object} &
    {\bf Date} & {\bf S('')} & {\bf Time(s)} & {\bf $\lambda_W$(\AA)}   \\ \hline
0023--26 & 10/7/94 & 1.8 & 2$\times$4$\times$600 & 2780---4080   \\ 
0035--02 & 20/10/95 & 1.7 &  4$\times$600  & 3010---4420 \\
& 22/10/95 &1.8 & 4$\times$600 & \\ 
0038+09 & 20/10/95 & 1.6 &  4$\times$600 & 3090---4540  \\  
        & 22/10/95 & 1.6 & 4$\times$600 &           \\  
0039--44 & 10/7/94 & 1.8 &  2$\times$4$\times$300 & 2730---4000  \\  
0105--16 & 11/7/94 & 1.7 & 2$\times$4$\times$600 & 2620---3850  \\ 
         & 21/10/95 & 1.7 & 4$\times$900 &            \\ 
0117--15 & 20/10/95 & 1.5 & 4$\times$900 & 2340---3440 \\
 & 21/10/95 & 1.4 & 4$\times$900 & \\ 
0235--19 & 20/10/95 & 1.5 & 4$\times$600 & 2270---3330  \\ 
         & 21/10/95 & 1.2 & 4$\times$600 &            \\
0252--71  & --  & -- & --    \\ 
0347+05  & 22/10/95 & 1.4 & 4$\times$600 & 2740---4030     \\
0409--75 & 22/10/95 & 1.4 & 4$\times$600  & 2170---3180 \\
1306--09 & 10/07/94 & 2.0 & 2$\times$4$\times$600 & 2500---3680  \\ 
         & 11/07/94 & 1.8 & 2$\times$4$\times$600 &             \\
1547--79 & 20/7/93  & 1.2 & 4$\times$900 &  2470---3630   \\
         & 21/7/93  & 1.7 & 4$\times$900 &    \\ 
1549--79 & 11/07/94 & 2.3 & 2$\times$4$\times$300 & 3190---4690  \\
1602+01  & 21/07/93 & 1.8 & 2$\times$4$\times$900 & 2510---3690   \\
1648+05  & --  & -- & --   \\
1932--46 & 10/7/94  & 2   & 2$\times$4$\times$600 & 2980---4380  \\
1934--63 & 23/7/93  & 1.7 & 4$\times$900 & 3100---4556  \\
1938--15 & 10/7/94  & 2   &  2$\times$4$\times$600 & 2530---3710 \\ 
         & 11/7/94  & 1.9 & 2$\times$4$\times$600 & \\
2135---20 & 23/7/93  & 1.3 & 4$\times$900 & 2245---3300  \\  
2211--17  & --  & -- & --    \\ 
2250---41 & 20/7/93  & 1.5 & 4$\times$900 & 2800---4110   \\
         & 21/7/93  & 1.4 & 4$\times$900 &    \\
         & 23/7/93  & 1.5 & 4$\times$900 &    \\ 
2314+03  & 20/10/95 & 1.7 & 4$\times$300 & 3010---4420  \\ 
\end{tabular}
\end{center}
\caption[Summary of the polarimetric  observations for the complete sample listed in Table 1.]
{Observational details of the B-band polarimetric observations for the objects in Table 1.
Column 3 lists the seeing FWHM for the observations. 
Column 4 lists the integration time for each complete cycle of four polarimetry measurements.
Column 5 lists the approximate, rest frame, wavelength range sampled by the B-band filter.}
\label{table:obs-pol-samp}
\end{table}

The B-band polarimetric observations were  obtained with EFOSC1 in polarimetric mode 
with a Wollaston prism and aperture mask in the beam. For 
the 1993 and 1994 runs the polarized signal was modulated by using the Cassegrain 
instrument rotator to rotate the field relative to the Wollaston prism successively through 
a sequence of four rotator positions separated by 45 degrees. However, for the 1995 run 
the signal was modulated by using a half-wave plate in the beam, and rotating the half-wave 
plate through the sequence 0,22.5,45,67.5 degrees. Details of the polarimetric observations are 
given in Table 3.

\subsection{Spectroscopic reductions}

The spectroscopic reductions followed the standard steps of bias subtraction, flat-fielding, 
wavelength calibration, atmospheric extinction correction, flux calibration and sky 
subtraction. The flux calibration was based on an average flux calibration curve derived 
from wide-slit observations of, typically, three flux calibration stars in each run. 
Comparisons between the flux calibration curves derived from individual standard star 
observations show that the relative flux calibration uncertainty is typically $\pm$5\% over 
most of the wavlength range, but rises to $\pm$10\% in the UV ($\lambda < 4200$\AA) and near-IR 
($\lambda > 8000$\AA). By using a relatively wide slit, making the observations as close as possible 
to the zenith, and, in some cases, making observations
with the slit aligned along the parallactic angle, the effects of differential refraction were minimised. 
Only in two cases does  
differential refraction significantly affect the accuracy of the 
flux calibration: PKS1547-79 (airmass $> 1.5$) and to a much
lesser extent PKS2314+03 (airmass $\sim 1.2$).

1 dimensional spectra of the near-nuclear regions
were extracted from the 2 dimensional frames using extraction
apertures that contained the bulk of
the visible continuum emission (typically 2 -- 4 arcseconds along the slit).  

\subsection{Polarimetry}

Following bias subtraction, cosmic ray removal, and correction for non-uniformities 
using flat-fields, the mean background level was determined for the `o'­ and `e'-ray 
images separately, using several apertures placed evenly around the source. The `o'­ and 
`e'-ray intensities for the radio galaxies were measured through  
circular apertures that included the bulk of the light in the near-nuclear regions
(typically 3 -- 6 arcseconds diameter) for each telescope rotator position or half-wave plate position, 
with the aperture size fixed through each cycle of rotator of four half-wave plate 
positions. The intensities for all the rotator positions were then combined according to the 
prescription of Tinbergen and Rutten (1992) to produce the final polarization degrees and 
position angles shown in Table 4. The advantage of this technique for measuring the 
polarization is that, since it involves the ratios of the `o'­ and `e'-ray intensities at 
each 
rotator/half-wave plate postion, it is not sensitive to small photometric variations between 
the images. By combining the intensity ratios from rotator positions separated by 90 
degrees, or half-wave plate positions separated by 45 degrees, any instrumental 
polarization produced in the instrument is automatically eliminated.

The uncertainties in the individual `o'- and `e'-ray intensities were estimated by 
combining the estimated uncertainty in the subtracted background (from the standard 
deviation in the background measurements), with the uncertainty due to the poissonian 
fluctuation  in the source+background counts in the source aperture. These uncertainties 
were then propagated through the calculation of the polarization degree and angle. The 
final polarization measurements and upper limits shown in Table 4 have been corrected 
for the positive bias in the polarization following the prescription of Simmonds and 
Stewart (1985).

\begin{table} 
\begin{center}
\begin{tabular}{lcc|cc}\\ \hline \hline
{\bf Object}  & {\bf P$_{meas}$} & {\bf P$_{int}$} & {\bf $\theta_{opt}$ ($^{\circ}$)} & {\bf $\theta_{radio}$ ($^{\circ}$)} \\ \hline
0023--26       &  $<$ 2.0        & $<$ 3.5      & - &  -    \\
0035--02       &  2.8 $\pm$ 0.3  & 5.6 $\pm$ 0.6  & 60$\pm$8 & 149$\pm$10  \\
0038+09       &    $<$ 1.9      & $<$ 2.5  & - & - \\
0039--44       &  4.8 $\pm$ 1.2  & 8.0 $\pm$ 2.0  & 177$\pm$4 & 99$\pm$9  \\
0105--16       &  $<$ 4.4        & $<$27          & - & -     \\
0117--15       &  8.2 $\pm$ 0.9  & 30.0$\pm$ 3.3  & - & -   \\
0235--19       &  $<$ 4.3        & $<$ 5.7     & - & -      \\
0252--71       &      -          & -           & - & -      \\
0347+05       &  $<$ 4.3        & $<$ 4.3      & - & -      \\
0409--75       &  $<$ 6.3        & $<$ 6.3     & - &  -     \\
1306--09       &  6.3 $\pm$ 1.3  &  8.4 $\pm$ 1.7  & 110$\pm$5 & 137$\pm$4 \\
1547--79       &  2.8 $\pm$ 1.0  &  2.8 $\pm$ 1.0  &  - & -  \\
1549--79       &  $<$ 1.8        &  $<$ 1.9       & - & -   \\
1602+01       &  $<$ 4.5        &  $<$ 7.5      & -  & -   \\
1648+05       &  -   & -      &  -  & - \\
1932--46       &  $<$ 1.6        &  $<$ 3.7  & - &	-        \\
1934--63       &  3.5 $\pm$ 0.5  &  8.5 $\pm$ 1.2 & 8.0$\pm$3.5  &  90$\pm$1  \\
1938--15       &  $<$ 5.5        &  $<$ 8.9      & - & -   \\
2135--20       &  $<$ 2.7        &  $<$ 3.6     & - & -    \\
2211--17       &  -              & -   &  - & - \\
2250--41       &  4.9 $\pm$ 0.7  &  18.6 $\pm$ 2.7  & 152$\pm$5 & 97$\pm$2  \\
2314+03       &  $<$ 1.2        &  $<$ 1.3       &  - & -   \\ \hline
\end{tabular}
\end{center}
\caption[B-band polarimetry of a complete sample of southern 2Jy radio galaxies.]
{Summary of the results of the B-band polarimetry of the sample of southern
radio galaxies corrected for postive bias (Simmons \& Stewart 1985).
P$_{meas}$ is the measured polarization corrected for bias, while  P$_{int}$ is the intrinsic polarization 
corrected for contamination by unpolarized narrow line emission, old stellar
populations and nebular continuum (see section 5.1 for details). The measured position
 angle of the polarization ($\theta _{opt}$)
is shown in Column~4 and, for comparision, the radio PA is included in Column~5.
The  polarization PA and radio PA  for 0035--02 are taken from Tadhunter et al. (1992),
the polarization and radio PAs for 1934--63 are from Tadhunter et al. (1994), and
for 2250--41 the polarization and radio PAs are taken from Shaw et al. (1995). }
\end{table}

For the 1993 and 1994 runs, which used the telescope rotator to modulate the 
polarization, the polarization angles were calibrated using observations of polarization 
standard stars observed using the same techniques in the same runs. However, for the 
1995 run, which used the half-wave plate, it was not possible to derive accurate 
polarization position angles because of problems with the initialisation of the half-wave 
plate at the end of each cycle; although the degrees of polarization measured for 
individual polarized sources and polarized standard stars were found to be consistent 
from one cycle to the next, large variations were found in the measured angles 
between the cycles. For the significantly
polarized objects observed in this run, the values of the polarization
listed in Table 4 represent the average of the polarization values measured
independently for each of the two cycles of half-wave plate positions. In the case of
objects without significant polarization but two cycles of observations from
the 1995 run, the data were analysed independently for each cycle, and
upper limits were derived from one cycle of observations only.   

As a final check on the reliability of the measurements, the individual `o'­ and `e'-ray 
intensity measurements for each cycle were checked to determine whether they followed 
the pattern expected for linearly polarized light. This allowed us to check for spurious 
polarizations which might arise, for example, from a cosmic ray affecting one of the
images. The degrees of 
polarization measured for the polarized standard stars were found to be consistent with 
the published values, within the estimated uncertainties. Apart from PKS1934-63 
(see Tadhunter et al. 1994), all the significant polarization measurements listed in 
Table 4 are based on two cycles of rotator/half-wave plate positions and have been 
checked for consistency between the cycles.


\section{Results}

With the exception of section 4.1, which will include results from the full
$z < 0.7$ sample of Tadhunter et al. (1993), the results presented below are for the
restricted sample ($0.15 < z < 0.7$) listed in Table 1.

\subsection{4000A break measurements}

Much of the early work on the continuum properties of radio galaxies concentrated on studies of
their broad-band colours. Although this work provided evidence for blue or UV excesses
in a large fraction of powerful radio galaxies compared with normal
ellipticals, the approach of using the broad-band colours has
the disadvantages that: (a) the broad-band filters may be contaminated in some redshift
ranges by emission lines; and (b) the sensitivity of a particular broad-band colour to the UV
excess depends on redshift. For example, the optical/near-IR broad-band colours  
are sensitive to UV excesses in high redshift ($z > 0.5$) radio galaxies 
(e.g. Lilly \& Longair 1984), but 
lose sensitivity for lower redshift objects.

\begin{figure*} 
\begin{center}
\epsfig{file=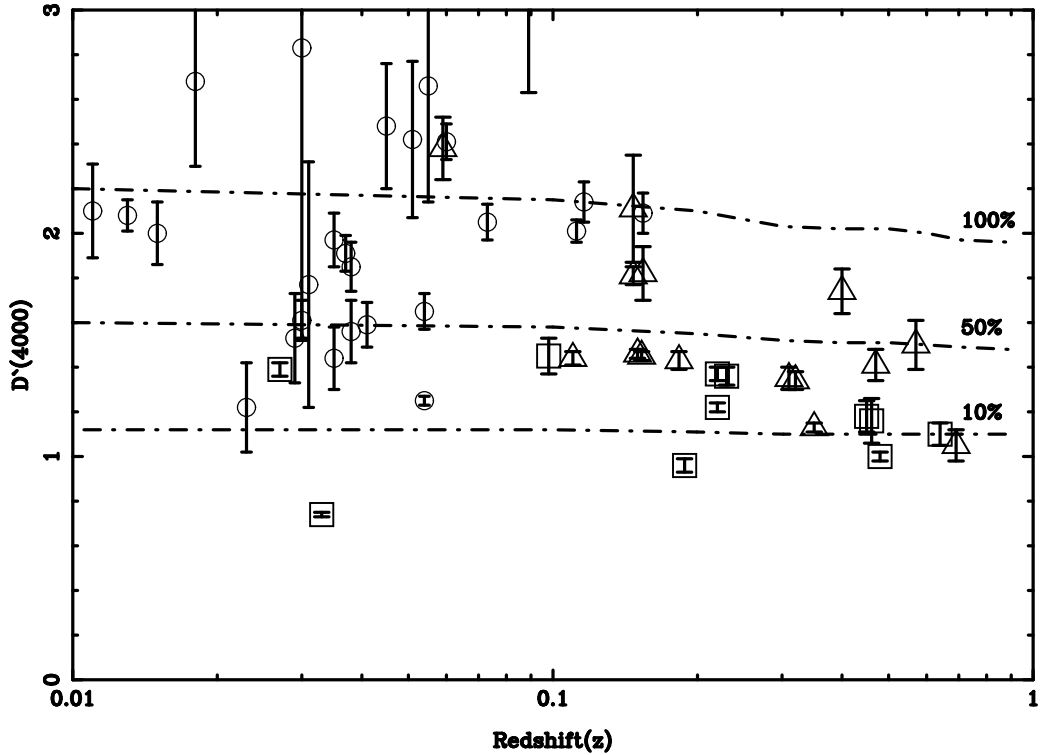,width=10cm,angle=270.0}
\end{center}
\caption[D(4000) versus redshift for a complete sample of radio galaxies.]
{The parameter D$^{\prime}$(4000) is sensitive to the presence of an excess UV flux compared to evolved stars below 4000\AA\
in the rest frame of an object. Two separate datasets have been plotted here: the low redshift (z $<$ 0.15) radio
galaxies with spectra
published in Tadhunter et al. (1993); and the complete sample of 0.15 $<$ z $<$ 0.7 radio galaxies
presented in Table 1. 
NLRG are shown as triangles, BLRG as squares, and WLRG as circles. The three solid curves represent 
the break of a passively evolving elliptical with a formation redshift, z$_f$=5. The models are from 
Bruzual \& Charlot (1993) and are based on an instantaneous burst with Salpeter (1955) initial mass function. The three
curves represent values of the break for
which 100, 50 and 10\% of the light below a rest frame wavelength of 4000\AA\ is due to evolved stars.}  
\label{figure:d4000}
\end{figure*}

An alternative way of quantifying the UV excess, which minimises these problems, is to 
use measurements of the
4000\AA\, continuum break, which is
prominent in the spectra of evolved stellar
populations (D(4000): Bruzual 1983). However, some care is required when measuring the 4000\AA\, 
break in powerful radio galaxies, since the bands used in the original definition of the D(4000) are contaminated
by [NeIII], [SII] and H$\delta$ emission lines, which can be strong in the spectra of radio galaxies. 
Therefore we use a modified version --- D$^{\prime}$(4000) --- which is defined as the ratio of the total flux in a bin 100\AA\, wide centred on
4200\AA\, (rest frame) to the total flux in a 100\AA\ wide bin centred on 3800\AA.  
The narrower bands used in this definition avoid strong emission lines, although the contamination by the higher
order Balmer lines may still be significant in some objects. The measurement 
of values of D$^{\prime}$(4000) 
significantly lower than this indicates a UV excess. It is worth emphasising that an additional advantage of using D$^{\prime}$(4000) to quantifiy the UV excess is that
it is relatively insensitive to flux calibration errors and intrinsic reddening effects. 

\begin{table} 
\begin{center}
\begin{tabular}{lcl} \hline \hline
{\bf Object} & {\bf D$^{\prime}$(4000)} & {\bf F$_{D4000}$}\\ \hline
0023---26 & 1.34 $\pm$ 0.04 & 0.63  \\
0035--02 & 1.22 $\pm$ 0.02 & 0.80\\
0038+09  & 0.97 $\pm$ 0.03  & 1.00 \\
0039--44 & 1.13 $\pm$ 0.02  & 0.87 \\
0105--16 & 1.74 $\pm$ 0.02 & 0.28 \\
0117--15 & 1.50 $\pm$ 0.11 & 0.50\\ 
0235--19 & 1.05 $\pm$ 0.15 & 0.95\\
0252--71 &   --  & -- \\ 
0347+05  & 1.13 $\pm$ 0.17 & 0.87 \\
0409--75 & 1.05 $\pm$ 0.07 & 0.95\\
1306--09 & 1.41 $\pm$ 0.07 & 0.60\\
1547--79 & 1.00 $\pm$ 0.02 & 1.00\\
1549--79 & 1.46 $\pm$ 0.02 & 0.59\\
1602+01  & 1.16 $\pm$ 0.10 & 0.84\\ 
1648+05 & 1.82 $\pm$ 0.12  & 0.26 \\
1932--46 & 1.36 $\pm$ 0.04 & 0.67\\
1934--63 & 1.43 $\pm$ 0.04 & 0.62\\
1938--15 & 1.18 $\pm$ 0.07 & 0.82\\
2135--20 & 1.10 $\pm$ 0.05 & 0.90\\
2211--17 & 1.54 $\pm$ 0.22 & 0.60 \\
2250--41 & 1.35 $\pm$ 0.05 & 0.66 \\
2314+03  & 1.37 $\pm$ 0.03 & 0.66 \\
\end{tabular} 
\end{center}
\caption{ Measurements of D$^{\prime}$(4000) for the complete sample of 2Jy radio galaxies.
The last column gives the estimated fraction of light
in the wavelength range 3750 --- 3950\AA\, that is {\it not}
associated with an old stellar population.}
\end{table}

Measurements of D$^{\prime}$(4000) for the  restricted $0.15 < z < 0.7$ sample in Table 1
are listed in Table 5 and plotted against redshift in Figure 1. For comparison purposes, 
values of D$^{\prime}$(4000) for the remainder of the $z < 0.7$ sample with spectra
published in Tadhunter et al. (1993)\footnote{
Note that the sample of low redshift objects with spectra published
in Tadhunter et al. (1993) is biased against objects 
with strong emission lines (NLRG,BLRG). This is because many of the strong emission
line sources had previously published spectra. Therefore they were not re-observed
as part of the survey, although they do, of course, form part
of the overall $z < 0.7$ complete sample of Tadhunter et al. (1993).} --- comprising   lower redshift objects ($z < 0.15$) --- are
also plotted in Figure 1. The measured values are compared with theoretical D$^{\prime}$(4000) 
predictions for passively evolving elliptical galaxies, as
derived from the models of Bruzual \& Charlot (1993). The final column of Table 5 gives an estimate of the
proportion of continuum light just below the 4000\AA\, break that is contributed by continuum components
other that an old stellar
population. This has been calculated from D$^{\prime}$(4000), under the assumption that the additional continuum
component has a flat spectrum.

There are several noteworthy features of these results:
\begin{itemize}
\item{\bf Higher redshift objects.} All the objects at $z > 0.15$ show evidence for
a UV excess, with the old stellar populations contributing $<$80\% of the flux
at wavelengths shortward of 4000\AA.
\item{\bf Lower redshift objects.} Although many lower redshift ($z < 0.15$) radio galaxies 
have D$^{\prime}$(4000) values consistent with a large fraction of light from an old stellar population, a significant
proportion of objects ($\sim$35\%) show smaller values, suggesting a significant UV excess
(see also Wills et al. 2002).
\item{\bf Weak line radio galaxies (WLRG).} Many of the lower redshift objects with evidence for a UV excess are
WLRG which do not show strong emission lines. This indicates that the presence of the UV excess is not always
linked to the classic signs of AGN activity.
\end{itemize}

Before considering the continuum shapes in more detail it is important to emphasise that other
factors than the UV excess can affect the 4000\AA\, break. For example, a lower metallicity than assumed in
the models can lead to lower values of D$^{\prime}$(4000), because of the reduced effect of metal line blanketing in
the atmospheres of the old stars at $\lambda < 4000$\AA. Although lower metallicity may be a factor in objects which
have D$^{\prime}$(4000) only marginally lower than the theoretical predictions in Figure 2, the metallicity would
have to be unusually low in order to explain many of the values  listed in Table 5.

\subsection{UV polarization measurements}

From the results of polarimetry measurements
presented in Table 4, it is clear that high UV polarization is rare
in our sample of intermediate redshift radio galaxies.
Only 7 out of 19 (37\%) of the objects in Table 1 
with UV polarization measurements are significantly
polarized in the UV; only 4 have measured polarization at the $P_B > 5$\% level; 
and none is highly polarized at the $P_B > 10$\%
level. For the majority of the sources it has only been possible to 
derive upper limits on the polarization.

It is notable that, of the significantly polarized sources, the two with the lowest UV polarization 
--- PKS0035-02 and PKS1547-79 --- are both broad line radio galaxies (BLRG), 
and in the case of PKS1547-79 it is possible that much of the UV polarization 
is due to dichroic extinction by dust 
in our Galaxy (Shaw et al. 1995). None of the remaining 7 BLRG in our sample is significantly 
polarized in the UV, despite the substantial UV excesses measured in all of these objects. Therefore 
the incidence of objects with  high UV polarization in the  population
of BLRG appears relatively low. In contrast, 
most of the objects with significant UV polarization are NLRG, and the
incidence of significant UV polarization amongst the NLRG is higher: 
5 out of 10 (50\%) of the NLRG with good polarization measurements show significant UV polarization.

The last two columns in Table 4 give the optical polarization (E-vector)  
and radio axis position angles, for the few cases for which it has been possible to measure the
PAs accurately. In most cases, the polarization angle is either close to the
perpendicular to the radio
axis (PKS0035-02, PKS0039-44, PKS1934-63) or the perpendicular to the UV structure axis 
(PKS2250-41: Shaw et al. 1995),
in line with measurements of other powerful radio galaxies (e.g. Cimatti et al. 1993). The 
only major exception is PKS1306-09, for which the polarization angle is closer to
the parallel to the radio axis. This suggests that the polarization mechanism in PKS1306-09 
may be different from that in the other sources.

\subsection{Nebular continuum}

The nebular continuum, which comprises a combination of 2-photon, free-free, and
free-bound continuum, as well as a pseudo continuum due to blended higher order Balmer lines (see
Dickson et al. 1995), is the best-constrained of the components which may contribute
to the UV excess. This is because it scales directly with the H$\beta$ flux and is relatively
insensitive to
the physical conditions, ionization and metallicity. The only major uncertainty with this component
is the line of sight reddening, which will act to reduce the flux of the observed nebular continuum
at UV wavelengths relative to the H$\beta$ flux.
\begin{table}
\begin{center}
\begin{tabular}{lc} \hline\hline
          & Nebular fraction \\ \hline
0023---26 & 0.37$\pm$0.04 \\
0035---02 & 0.11$\pm$0.02 \\
0038+09 & 0.05$\pm$0.01  \\
0039---44 & 0.19$\pm$0.02 \\
0105---16 & 0.12$\pm$0.02 \\
0117---15 & 0.29$\pm$0.06 \\
0235---19 & 0.11$\pm$0.02 \\
0252---71 & -- \\
0347+05  &  -- \\
0409---75 & 0.12$\pm$0.04 \\
1306---09 & 0.10$\pm$0.01 \\
1547---79 & -- \\ 
1549---79 & 0.14$\pm$0.02 \\
1602+01  &  0.17$\pm$0.08 \\
1648+05  & 0.11$\pm$0.03 \\
1932---46 & 0.18$\pm$0.02 \\
1934---63 &  0.15$\pm$0.03 \\
1938---15 &  0.15$\pm$0.02 \\
2135---20 & 0.14$\pm$0.02 \\
2211---17 & 0.05$\pm$0.01 \\
2250---41 & 0.27$\pm$0.02 \\
2314+03  &  0.03$\pm$0.01 \\ \hline
\end{tabular}
\end{center}
\caption[The nuclear nebular fraction]
{The fractional contribution of the nebular continuum to the UV flux. The ratios have been
calculated from the mean fluxes in a 100\AA\ bin centred on 3590\AA\ (rest-frame), 
for both 
the observed nuclear spectra and the theoretical nebular spectra.
In each case the nebular spectrum was calculated from the H$\beta$ flux 
assuming an electron density of 100cm$^{-3}$ and a temperature of 15000~K and zero reddening.}
\label{table:neb-fraction}
\end{table}

Prior to the more detailed continuum modelling described in section 4.4, the nebular continuum spectrum was
calculated from the measured H$\beta$ flux for each object using the STARLINK DIPSO 
package. The spectrum of the pseudo continuum due to blended higher order Balmer lines was also generated
(see Dickson et al. 1995) and added to the DIPSO output. The calculation of the nebular continuum was
made for an electron temperature of 15,000~K and a density of 100 cm$^{-3}$.

The estimated fractional contribution of the nebular continuum to the observed UV continuum 
of all the objects in Table 1 with good H$\beta$ and UV continuum flux estimates is shown in Table 6. 
These estimates were made for a 100\AA\, continuum bin centred just below the Balmer continuum recombination
break --- where the nebular continuum has its peak flux. Since no account has been taken of reddening, these
estimates represent an upper limit on the true fractional contribution of the nebular component to the UV
continuum. The effect of the reddening of the nebular continuum on the continuum modelling will be 
discussed in section 4.4 below.

It is clear from Table 6 that the nebular continuum makes a significant contribution to the UV continuum
in all of the sample objects, comprising 3 --- 40\% of the continuum below the Balmer break. 

\subsection{Modelling}

In order to gain further information about the nature of the UV continuum 
we have attempted to model the continuum
spectral energy distributions (SEDs) in terms of various spectral components which may, 
potentially, contribute to the UV excess. Prior to this modelling effort we 
subtracted the nebular continuum calculated as described
in section 4.3, and assuming no reddening in this component.

\begin{figure*}
\epsfig{file=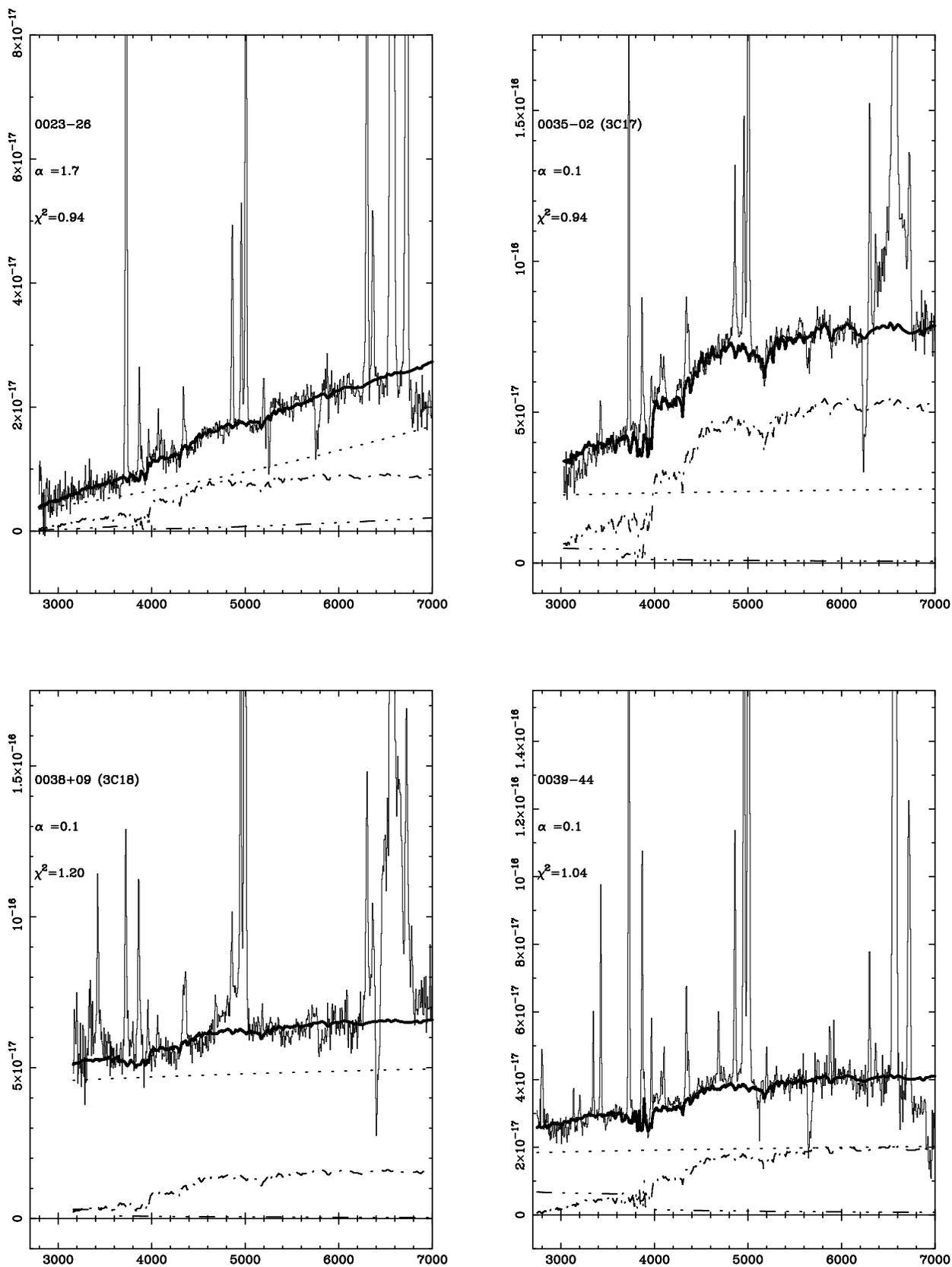,width=16.5cm,height=22cm}
\caption[Two component continuum  models of the  radio galaxy sample]
{Power-law plus 15Gyr elliptical galaxy fits to the continua of the complete sample of radio galaxies. The best-fit  model 
continuum spectrum is  indicated by the thick continuous line.  
The 15Gyr elliptical galaxy component is indicated by a dot-dash line, the power-law component
by a dotted line and the nebular continuum by a dot-dot-dot-dash line. The fluxes are in units 
of ergs cm$^{-2}$ s$^{-1}$ \AA$^{-1}$.}
\label{figure:sample-model}
\end{figure*}
 
\begin{figure*}
\setcounter{figure}{1}
\epsfig{file=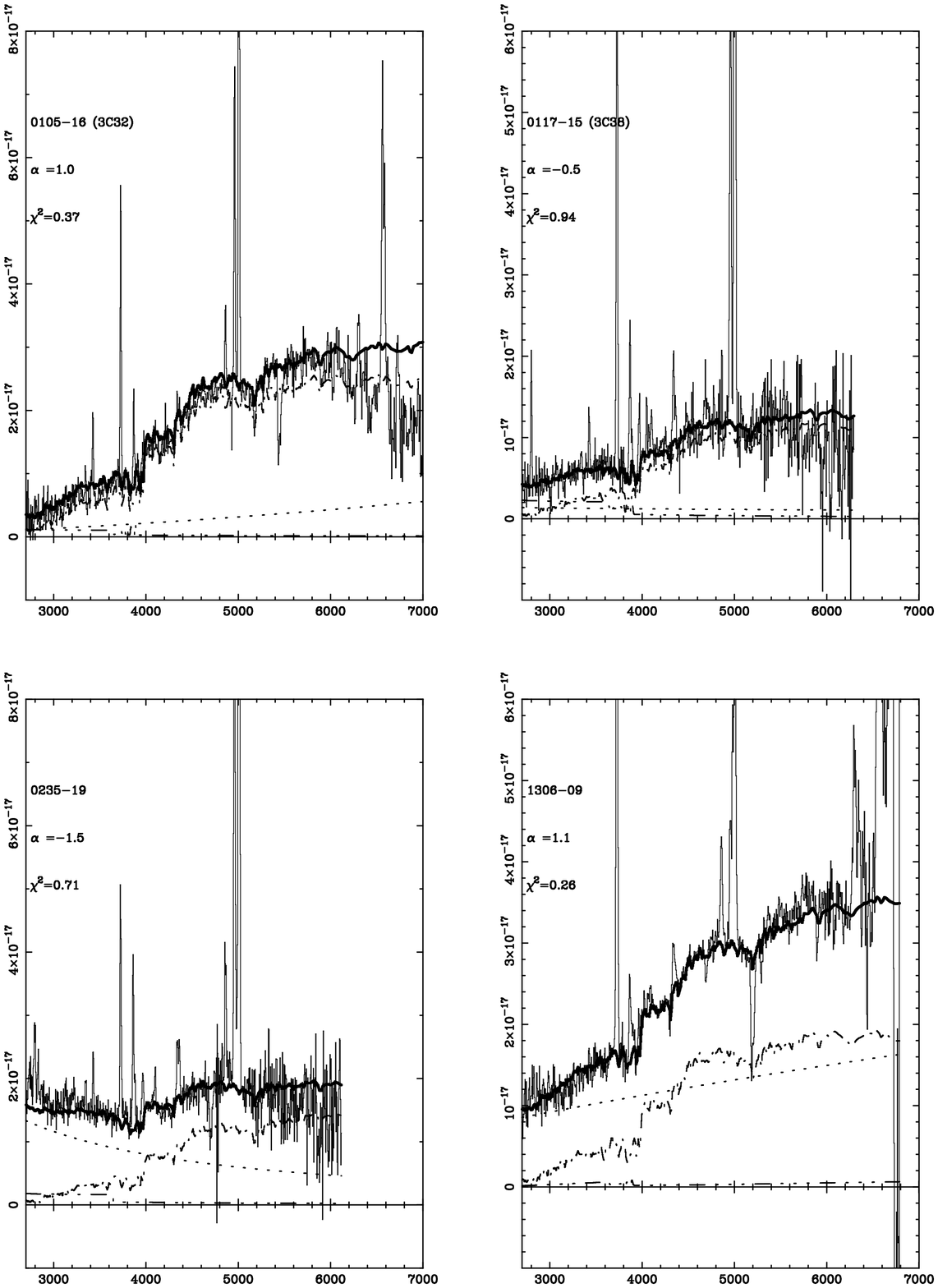,width=16.5cm,height=22cm}
\caption{Cont.}
\end{figure*}
 
\begin{figure*}
\setcounter{figure}{1}
\epsfig{file=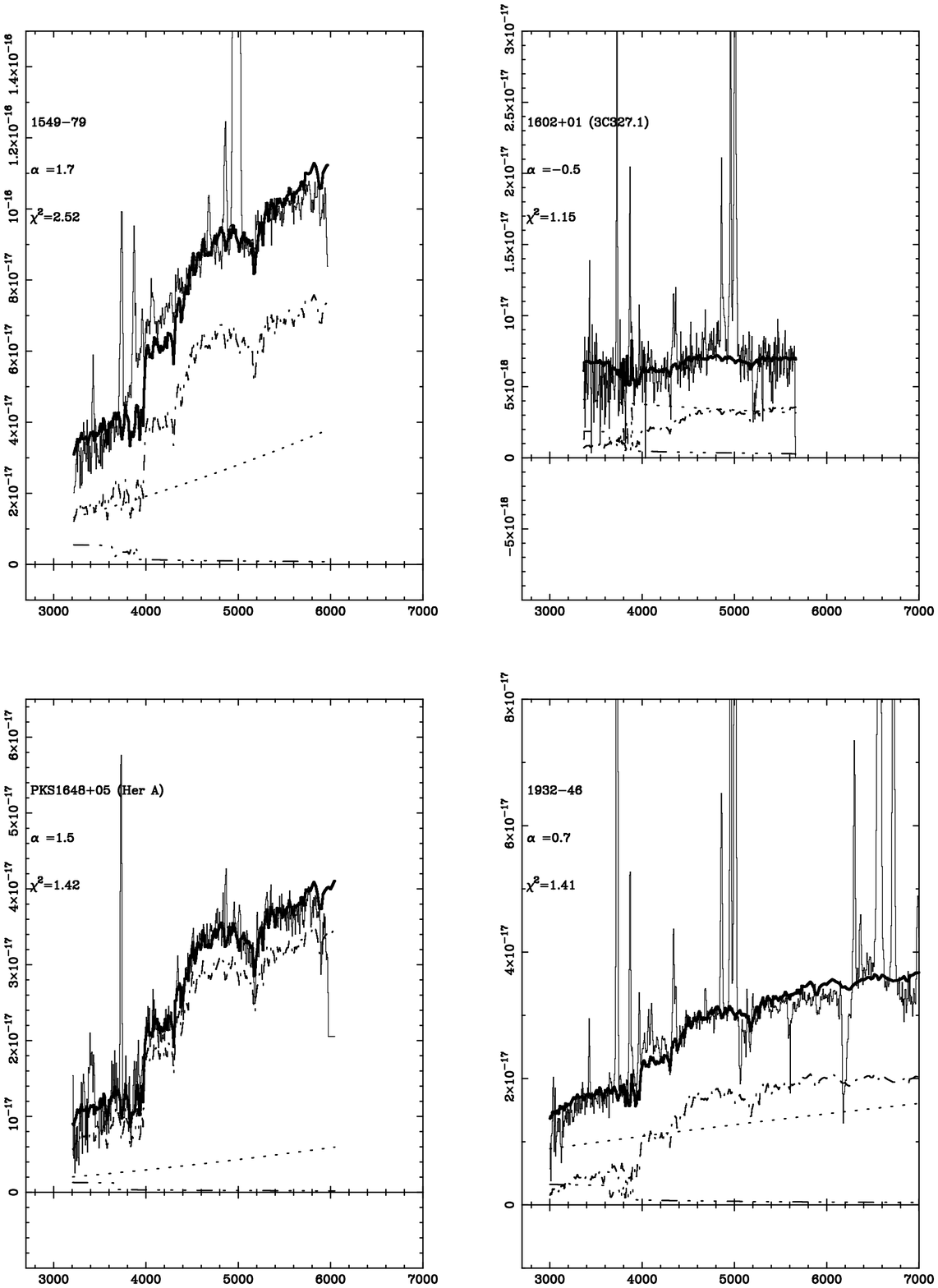,width=16.5cm,height=22cm}
\caption{Cont.}
\end{figure*}
 
\begin{figure*}
\setcounter{figure}{1}
\epsfig{file=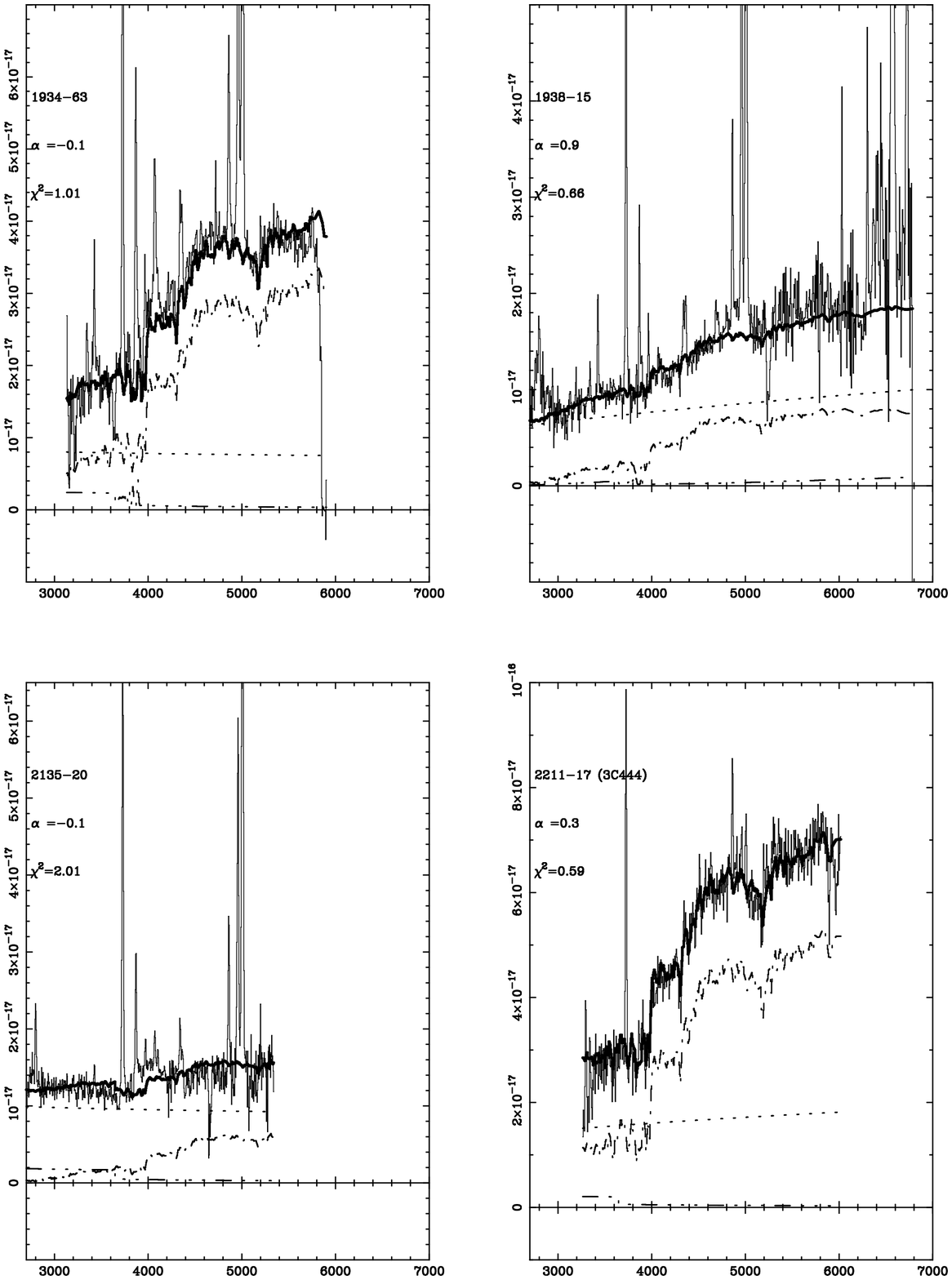,width=16.5cm,height=22cm}
\caption{Cont.}
\end{figure*}

Given that radio galaxies have early-type continuum morphologies at optical 
wavelengths, we started by assuming that the nebular-subtracted optical-UV continuum 
SEDs comprise a combination of an old stellar population and a power-law. For the old 
stellar population we used the instantaneous burst model of Bruzual and Charlot (1983), 
for a Salpeter IMF and age (since the starburst) of 15Gyr. The power-law was used to 
represent the ``active'' component e.g. scattered AGN light or direct AGN light, 
with the power-law spectral index allowed to vary over the range $-6 < \alpha 
< +6$ ($F_{\lambda} \propto \lambda^{+\alpha}$).

As a first step to modelling the spectra, continuum fluxes were measured 
in a number of wavelength bins, chosen to avoid strong emission lines, cosmetic 
defects and regions of poor sky subtraction or poor flux calibration. Errors for the 
continuum flux measurements were calculated by quadratically combining the Poisson
noise in sky+object signal, an assumed relative flux calibration error of $\pm$5\%, 
and an estimate of the systematic error in the background subtraction, determined by
examining spatial slices covering the wavelength range of each bin, as extracted 
from the sky-subtracted 2D spectra. The fluxes and errors were estimated in 15 -- 30 
bins for each object (depending on the useful wavelength range for each spectrum), and 
the fitting was then performed using the minimum $\chi^2$
technique detailed in Tadhunter et al. (1996).

\begin{table} 
\begin{center}
\begin{tabular}{lccc} \hline \hline
{\bf Object}    & {\bf $\chi^2$}  &  {\bf $\alpha$}         
&{\bf F$_{model}$}\\ \hline
0023--26        &    2.07         & 3.3                      & 0.55 \\
0035--02        &    0.94         & 0.1$^{+0.6}_{-1.4}$    
& 0.60 $^{+0.16}_{-0.14}$  \\
0038+09         &    1.20     & 0.1$^{+0.4}_{-1.8}$    
& 0.91$^{+0.09}_{-0.45}$  \\
0039--44        &    1.04         & 0.1$^{+0.6}_{-1.0}$    
& 0.70$^{+0.12}_{-0.16}$  \\
0105--16        &    0.69         & 2.7$^{+4.2}_{-3.8}$   
& 0.25$^{+0.36}_{-0.24}$   \\
0117--15        &    0.95         & -0.5$^{+2.4}_{-6.8}$   
& 0.24$^{0.23}_{-0.16}$   \\
0235--19        &    0.71          & -1.5$^{+1.2}_{-1.0}$  
& 0.67$^{+0.10}_{-0.08}$     \\
0347+05         &    3.10         & -0.9                  &0.88   \\
0409--75        &    4.30         & 1.7                  &0.91    \\
1306--09        &    0.26         & 1.1$^{+0.8}_{-2.6}$     
&0.64$^{+0.23}_{-0.24}$  \\
1549--79        &    2.52         & 1.7                     & 0.45 \\
1602+01         &    1.15         & -0.5$^{+1.4}_{-1.4}$    
&0.69$^{0.24}_{-0.23}$   \\
1648+05         &    1.42         & 1.7$^{+1.4}_{-0.8}$     
&0.24$^{+0.17}_{-0.05}$  \\
1932--46        &    1.41         & 0.7$^{+0.8}_{-1.8}$     
&0.62$^{+0.16}_{-0.14}$  \\
1934--63        &    1.01         & -0.1$^{+1.8}_{-1.8}$   
&0.45$^{+0.30}_{-0.23}$    \\
1938--15        &    0.66         & 0.9$^{+0.8}_{-1.8}$    
&0.70$^{+0.17}_{-0.32}$    \\
2135--20        &    2.01         & -0.1                    &0.82  \\
2211--17        &    0.59         & 0.3$^{+1.4}_{-2.2}$   
&0.54$^{+0.30}_{-0.23}$   \\
2250--41        &    0.69         & 0.9$^{+0.6}_{-2.8}$      
& 0.55$^{^0.35}_{-0.22}$ \\
2314+03         &    3.96         & -0.3                   
&0.82   \\ \hline
\end{tabular}
\end{center}
\caption[Two component fit to the southern sample]
{Results of modelling the SED's of the  southern sample of radio galaxies  
 using two-component --- 15 Gyr galaxy plus power
law --- models. Column~2 is the reduced $\chi^2$. Column~3 is the spectral index of the power-law 
(f$_{\lambda} \propto \lambda^{\alpha}$). Column~4 lists the fraction of the total model continuum flux (including the nebular component) contributed by the fitted power-law component in a wavelength
bin just shortward of the 4000\AA\, break (3750 --- 3850\AA\,). Where no confidence interval is 
quoted the probability
of the best--fit model was less than 5\%. }
\label{table:sample-model}
\end{table}

The results of the modelling are shown in Table 7 and Figure 2. Only objects
with significant fits are shown in the Figure. In most objects the power-law$+$E-galaxy 
model provides an adequate fit to the data, with the power-law component contributing 
20 --- 90\% of the continuum just below the 4000\AA\, break (see final column in Table 7). 
This confirms the results based on the 4000\AA\,
break measurements described in section 4.1. Note, however, that the proportional contribution of 
the power-law to the UV continuum at wavelengths just below 4000\AA\, is systematically less than estimates 
derived from  D$^{\prime}$(4000) (compare the last columns of Tables 5 and 7). 
This difference is likely 
to be due to the fact that the  D$^{\prime}$(4000)
measurements were made before the subtraction of the nebular continuum, and therefore 
over-estimate the contributions of the putative power-law
components.  No modelling was performed for
PKS1547-79 because of the effects of differential atmospheric refraction
on the flux measurements. Also, poor sky subtraction, weak signal and relatively
short exposure times are likely to be responsible for the
relatively poor fits obtained for PKS0409-75 and PKS0347+05. 

\begin{figure}
\setcounter{figure}{1}
\epsfig{file=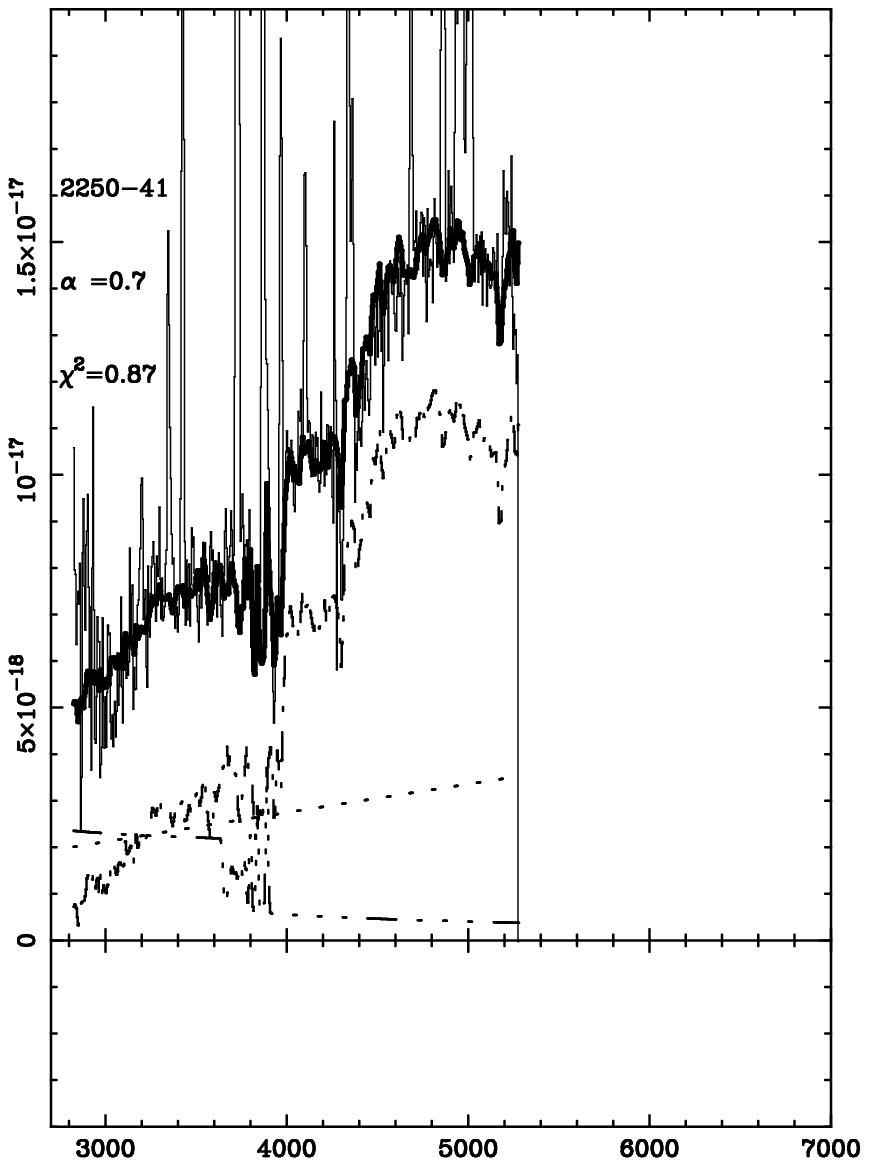,width=8.0cm,height=10.5cm}
\caption{Cont.}
\end{figure}

One interesting general feature of the results in Table 7 is that, 
in the majority of cases, the fitted power-law spectral indices are redder 
than the mean $\alpha = -1.3$ derived for the optical/UV continua of quasars
by Christiani \& Vio (1990) over the range 1000 -- 5500\AA. They are
also redder than most $z < 0.7$ quasars in the
southern 2Jy sample of Tadhunter et al. (1993), which have measured optical 
(3000 --- 5500 \AA) spectral 
slopes in the range $-2.8 < \alpha < -0.5$; and high redshift quasars in the sample
of Kuhn et al. (2001) which have UV-optical slopes in the range $-2.2 < \alpha < -0.8$. 
Therefore, if the power-law component represents direct or scattered AGN light, then 
it is likely that this component is significantly reddened in many of the objects.

Given that the results of the modelling are potentially sensitive to the effects of 
reddening on the subtracted nebular continuum\footnote{Note that the effects
of reddening on the power-law component will be reflected in the fitted spectral 
index for that component. We assume in our modelling that the 15Gyr old stellar population is not 
significantly affected by reddening.}, we repeated the modelling following subtraction of
a nebular continuum reddened by the amount determined from the emission line Balmer decrement assuming
case B recombination theory, 
for cases in which the redenning was deemed significant. The results are 
shown in Table 8. Subtraction of the reddened nebular continuum leads to significantly 
improved fits in the case of PKS0023-26 and PKS1648+05, but in most other cases the 
improvement is marginal. As expected, given that less nebular continuum is being 
subtracted from the UV excess, the proportional contribution of the power-law component 
is larger in these models, and the
power-law slope is bluer.

\begin{table} 
\begin{center}
\begin{tabular}{lcccc} \hline \hline
{\bf Object}    & {\bf $\chi^2$}  &  {\bf $\alpha$}       
&{\bf F$_{model}$} & {\bf E(B-V)} \\ \hline
0023--26        &    0.94         & 1.7$^{+0.2}_{-1.0}$ & 0.66$^{+0.18}_{-0.28}$    & 1.18    \\
0105--16        &    0.37         & 1.9$^{+4.2}_{-3.8}$ & 0.24$^{+0.35}_{-0.23}$    & 1.10    \\
1306--09        &    0.12         & 0.7$^{+1.0}_{-2.6}$ & 0.67$^{+0.25}_{-0.27}$     & 0.80   \\
1549--79        &    1.59         & 1.5$^{+0.2}_{-0.8}$ & 0.58$^{+0.08}_{-0.16}$    & 0.61    \\
1648+05        &    0.95         & 1.5$^{+1.0}_{-4.4}$ & 0.27$^{+0.11}_{-0.06}$    & 0.94    \\
1938--15        &    0.86         & 0.5$^{+0.8}_{-2.4}$ & 0.75$^{+0.21}_{-0.29}$     & 0.95   \\ \hline
\end{tabular}
\end{center}
\caption[Two component fit with reddening to the southern sample]
{Results of modelling the SEDs of the  southern sample of radio galaxies using two--component --- 15 Gyr galaxy plus power
law --- models including reddening of the nebular continuum. The columns are the same as  Table~7 except 
column~5, which gives the degree of reddening estimated from the Balmer decrement.}
\end{table}

However, for three objects --- PKS1549-79, PKS2135-20, PKS2314+03 (3C459) --- we failed 
to obtain good fits to the nebular continuum subtracted SEDs using two component
E-galaxy$+$power-law models, despite the good quality of the
data. In all of these cases we found a significant excess of the
data over the models in the wavelength range 3750 --- 4300\AA. In addition, higher order
Balmer absorption lines are detected in all three objects either in our
own spectra  or published spectra (see di Serego Alighieri et al. 1997
for a higher resolution spectrum of PKS1549-79), and both PKS2135-20 and  PKS2314+03
have relatively blue continuum slopes at wavelengths larger than 4500\AA. 
All of these features are consistent with a significant contribution from
young stellar populations (see Tadhunter et al. 1996, Robinson et al. 2000). 
Therefore we have attempted to model the continua of the three
objects with various combinations of old and young  stellar
populations (using the instantaneous burst models of Bruzual \& Charlot 1993) and power-laws. 
We have also tried models in which the young stellar populations
were reddened by varying degrees. The best fitting models
are shown in Table 9, and compared with the data in Figure 3. In the case of PKS1549-79
and PKS2135-20 we obtained an adequate fit using models which combine an old (15Gyr) and
young stellar populations (reddened in the case of PKS1549-79). On the other hand,
in the case of PKS2314+03 the best fit comprised a combination of young stellar populations
with two starburst ages plus a power-law. Given that the spectrum of
PKS2135-20 has a relatively low S/N, that of PKS1549-79 has a relatively small spectral
coverage, and that of PKS2314+03 may be affected by differential
atmospheric refraction effects (see section 3.1), the modelling estimates of the ages of the 
young stellar populations
should not be regarded definitive. Nonetheless, these results provide strong
evidence that the optical/UV continua of all three objects
are dominated by the light of young stellar populations with ages in the range 0.1 -- 2Gyr.

\begin{table}
\begin{center}
\begin{tabular}{lccccc}  \hline \hline
{\bf Object} & {\bf $\chi^2$}      & {\bf Age }  & {\bf F$_{ys}$} &{\bf F$_{pl}$} & {\bf E(B-V)} \\ \hline
PKS1547--79  & 0.56          & 1.0          & 0.88 & -- & 0.4 \\
PKS2135--20  & 0.74          & 0.1         & 0.90 & -- & 0.0 \\ 
PKS2314+03   & 0.29          & 1.0,0.5          &0.34,0.28 &0.36 & 0.0 \\
             & 0.51          & 2.0,0.1       &0.64,0.38 & -- & 0.0,0.2 \\ \hline
\end{tabular}
\end{center}
\caption[Young stars in PKS1549--79 and PKS2135--20]
{Best fitting models for PKS1549--79, PKS2135--20 and PKS2314+03 that include young 
stellar populations, yield formally significant fits, and reproduce the excess flux
seen in the wavelength interval 3900\AA\---4300\AA\ over the  two component (power--law$+$E-galaxy fits)
models. In the case of PKS2314+03 the fitting results
are shown for models both with and without a power-law component. The third column gives
the age in Gyr for the best-fitting young stellar populations, while
the fourth and fifth columns give the fractional contributions
of the young stellar populations and power-law to the total flux in the normalising
bin respectively.}
\label{table:young stars}
\end{table}

\begin{figure*}
\begin{center}
\epsfig{file=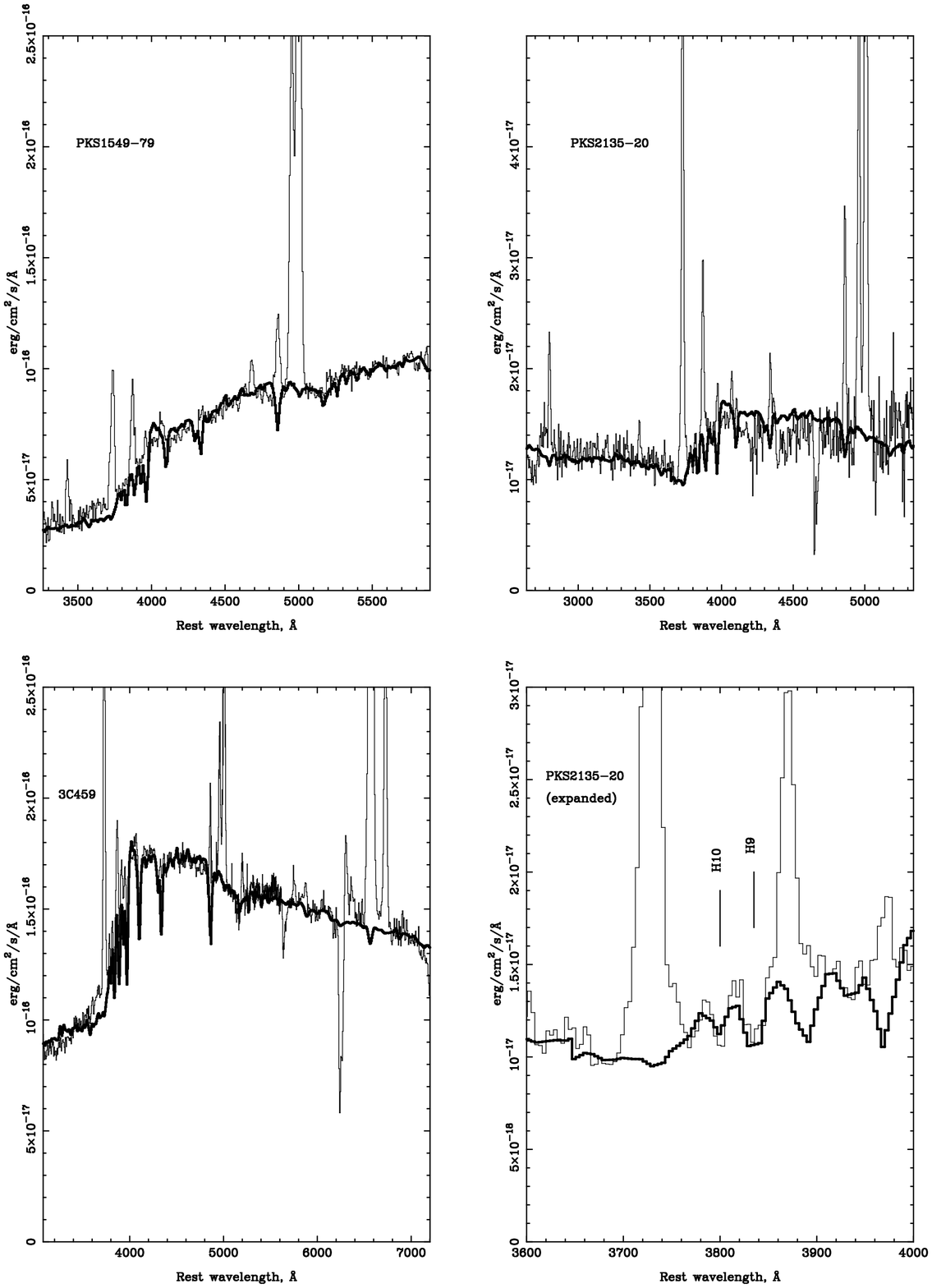,width=14.5cm,height=19.5cm}
\end{center}
\caption[Young stars in PKS1549--79, 3C459 and PKS2135--20]
{Fits to the continua of PKS1549--79, PKS2135--20 and  PKS2314+03(3C459) that include young stellar
populations. While two--component elliptical galaxy plus power--law or quasar models cannot reproduce 
the continuum SEDs of PKS1549--79,  PKS2135--20 and PKS2314+03(3C459), models in 
which the spectra are dominated by young stars 
are successful. PKS1549--79 has been modelled by the spectrum of an  1 Gyr population reddened by
E(B-V)=0.4 while PKS2135--20 is shown plotted with a model comprising an unreddened 0.1 Gyr population,
and PKS2314+03 (3C459) has been modelled with a combination of a 1Gyr starburst, a 0.5Gyr starburst
and a power-law (see Table 9). The lower right hand panel shows an expanded plot of the region
around the Balmer break in PKS2135--20 with the higher Balmer lines (H9,H10) labelled. In
all cases the best fitting continuum models are indicated by thick solid
lines.}
\label{figure:1549-2135-ystars}
\end{figure*}

%

\subsection{Notes on individual objects}

{\bf PKS0023-26.} This CSS radio galaxy has a large UV excess, but a low polarization, and no sign of
broad permitted lines in its spectrum. A possible starburst candidate.
\vglue 0.2cm\noindent
{\bf PKS0035-02 (3C17).} The strength of the broad permitted lines
in this BLRG suggests that much of its UV excess is due to direct AGN light. However,
given the significant polarization detected at both UV wavelengths 
(this paper) and optical wavelengths (Tadhunter et al. 1992), scattered AGN light may also contribute. Optical
synchrotron emission is a plausible alternative to scattered AGN light in this source,
given the distorted S-shaped radio structure, and relatively strong core-jet visible
in high resolution radio maps (Morganti et al. 1999).
\vglue 0.2cm\noindent
{\bf PKS0038+09 (3C18).} Direct AGN light is likely to contribute much of the UV excess in this
BLRG which has a low UV polarization.
\vglue 0.2cm\noindent
{\bf PKS0039-44.} This FRII radio galaxy has a large UV excess, significant UV polarization, but no
broad permitted lines. Although the scattered AGN light is likely to be significant, given the
relatively low level of intrinsic polarization (Table 4), 
it is unlikely that this component dominates the
UV continuum. A possible starburst candidate.
\vglue 0.2cm\noindent
{\bf PKS0105-44 (3C32).} This FRII radio galaxy has a modest UV excess, no significant UV polarization, and
no clear detection of broad permitted lines. The origin of its UV excess is not clear. 
\vglue 0.2cm\noindent
{\bf PKS0117-15 (3C38).} This high redshift FRII radio galaxy is one of the most highly polarized
objects in our sample. Given the high intrinsic polarization (Table 4), it is likely that scattered AGN light
dominates the UV excess in this source.
\vglue 0.2cm\noindent
{\bf PKS0235-19.} This FRII radio galaxy shows a large UV excess, but has a low UV polarization.
Weak broad wings are marginally detected to the H$\beta$ and MgII(2800) permitted lines, but
deeper spectroscopic observations are required to confirm that this is a BLRG in which the
UV excess is dominated by direct AGN light.
\vglue 0.2cm\noindent
{\bf PKS0252-71.} Contamination of images and spectra by a forground galaxy preclude detailed study
of the optical continuum in this CSS radio galaxy.
\vglue 0.2cm\noindent
{\bf PKS0347-05.} This unusual FRII radio galaxy has broad permitted lines but only weak narrow lines
(di Serego Alighieri et al. 1994).
Although it has not been possible to model the continuum of this object in detail because of
a poor sky subtraction, given the low level of UV polarization, it is likely that direct AGN light
contributes much of its UV excess.
\vglue 0.2cm\noindent
{\bf PKS0409-75.} This FRII radio galaxy is the highest redshift source in our sample. This
object clearly has a large UV excess, but it has not been possible to model its continuum in
any detail because of a poor sky subtraction. The relatively low UV polarization and failure
to detect broad permitted lines make this a strong starburst candidate.
\vglue 0.2cm\noindent
{\bf PKS1306-09.} This CSS radio galaxy has a large UV excess, only narrow emission lines, and
significant UV polarization. Given that the polarization E-vector angle is closer to the parallel
than the perpendicular to the radio axis, the polarization mechnaism may not be anisotropic
scattering, but some alternative mechanism such as dichroic absorption or synchrotron emission. The
low level of intrinsic polarization suggests that, in the case that the polarization mechanism
is anisotropic scattering, the scattered component does not dominate the UV excess.
\vglue 0.2cm\noindent
{\bf PKS1547-79.} This BLRG is one of the most luminous sources in our sample.
Its continuum properties have been extensively
discussed in Shaw et al. (1995), who argue that at least some of the significant UV
polarization measured in this source may be due to dichroic absorption in our Galaxy. Differential
diffraction effects preclude detailed modelling of its continuum, but it is likely that the UV
excess in this object is dominated by direct AGN light.
\vglue 0.2cm\noindent
{\bf PKS1549-79.} The narrow emission line properties of this extraordinary flat spectrum
radio galaxy have been discussed in detail by Tadhunter et al. (2001). It is likely
that this is an intrinsically compact radio source in which the radio jets
are pointing close to our line of sight, but the quasar nucleus is entirely
extinguished by dust on a kpc-scale. No broad permitted lines
are detected in our optical spectra of this source, and we  detect no significant
UV polarization in our observations. However, di Serego Alighieri et al. (1997) report 
significant optical
polarization at the $\sim$3\% level in both the optical continuum and 
[OIII] emission lines in their
smaller aperture observations. This polarization could have either a scattering or a dichroic
origin, however, it is unlikely that the polarized component dominates the UV excess. Both our
continuum modelling, and the detection of Balmer lines in absorption in the higher resolution
observation of di Serego Alighieri et al. (1997) are consistent with a starburst dominating
the optical/UV continuum in this source.
\vglue 0.2cm\noindent
{\bf PKS1602+01 (3C327.1).} This  FRII radio galaxy has been discussed in detail by Shaw et al. (1995), who classify
it as a BLRG on the basis of the clear detection of broad H$\beta$ emission lines. Given the 
failure to detect significant UV polarization it is likely that the UV excess 
in this source is entirely dominated by
direct AGN light. 
\vglue 0.2cm\noindent
{\bf PKS1648+05.} This powerful radio source, with a morphology between FRI and FRII, has weak
narrow emission lines in its optical spectrum, and the smallest UV excess of any source in
Table 1. The continuum measurements presented in this paper refer to the brighter of the two nuclei
visible in optical images of the core region. The cause of the apparent UV excess is unknown.
\vglue 0.2cm\noindent
{\bf PKS1932-46.} Villar-Martin et al. (1998) report the detection of a broad H$\alpha$ emission line
in this FRII radio galaxy (also visible in the spectrum in Figure 2), 
but it shows no significant UV polarization. The residuals from the
two component continuum model fit suggest a significant contribution from a starburst
component, in addition to direct AGN light. The spectacular extended emission line
nebulosity surrounding the galaxy is described in detail by Villar-Martin et al. (1998).
\vglue 0.2cm\noindent
{\bf PKS1934-63.} As discussed in Tadhunter et al. (1994), this GPS radio galaxy is significantly
polarized in the UV with the polarization E-vector close to perpendicular to the radio axis. No
broad permitted lines are detected in our optical spectra. It is likely that scattered AGN
light makes a significant contribution to the UV excess in this source.
\vglue 0.2cm\noindent
{\bf PKS1938-15.} This FRII shows broad H$\beta$ and MgII(2800) emission lines in its optical/UV
spectrum, but is not significantly polarized in the UV. Therefore its UV excess is likely to be dominated
by direct AGN light.
\vglue 0.2cm\noindent
{\bf PKS2135-20.} Shaw et al. (1995) have reported the detection of broad MgII(2800) in this CSS radio
galaxy. However, its classification as a BLRG is not certain because of the
possibility that blended narrow high ionization lines make a contribution
to the broad wings of MgII(2800) in our low resolution spectrum
(e.g. Dey \& Spinrad 1996, Tran et al. 1998). The detection of the  Balmer break
higher Balmer absorption lines (H9, H10: see Figure 3) in our optical spectrum suggests a dominant contribution
from a starburst component, and this is confirmed by our continuum modelling. 
PKS2135-20 may be a radio-loud example
of the ``post-starburst quasar'' phenomenon first reported by Brotherton et al. (1999).
\vglue 0.2cm\noindent
{\bf PKS2211-17.} Despite the fact that it belongs to the class of
weak line radio galaxies, with no broad permitted lines detected in its optical spectrum, this
FRII radio galaxy displays a substantial UV excess. A significant contribution from a 
starburst component is suspected. 
\vglue 0.2cm\noindent
{\bf PKS2250-41.} The continuum properties of this FRII radio galaxy have been discussed in
detail by Shaw et al. (1995) and Dickson et al. (1995). Its optical spectrum shows only
narrow emission lines, but it is significantly polarized in the UV with the polarization
E-vector close to the perpendicular to the UV structural axis. The relatively high
instrinsic polarization suggests that scattered AGN light makes a substantial
contribution to the UV excess in the nuclear regions. Clark et al. (1997) and Villar-Martin et al. (1999)
describe detailed spectroscopic and imaging observations of a spectacular jet-cloud
interaction on the west side of the nucleus.
\vglue 0.2cm\noindent
{\bf PKS2314+05 (3C459).} The optical/UV spectrum of this FRII radio galaxy is clearly dominanted
by the light of a young stellar population, with the Balmer break and higher Balmer absorption lines
clearly detected (see also Miller 1981). Our spectrum shows no clear evidence 
for broad permitted lines.

\section{Discussion}
\subsection{The scattered AGN component}

Our polarization measurements of a complete, optically unbiased sample allow us to 
gauge for the first time the significance of the scattered AGN component in UV 
continua of the general population of powerful radio galaxies. 

We detect significant UV 
polarization  in only a relatively small fraction ($\sim$37\%) of our sample, and the level of 
polarization is relatively low in the objects that are polarized ($P_B < 10$\%). Moreover, 
scattering of the anisotropic quasar light is not the only polarization mechanism, but other 
mechanisms such as dichroic absorption and synchrotron radiation may also contribute. 
Therefore it is safe to conclude that the scattered AGN component does not dominate 
the UV light in the majority of sources in our sample. For pure scattered light we would 
expect to measure polarization  at the $P > 10$\% level, even allowing for the geometrical 
dilution caused by integrating over the broad radiation cones of the illuminating AGN
(Young et al. 1995, Manzini \& di Serego Alighieri 1996). 

\begin{figure*}
\begin{center}
\epsfig{file=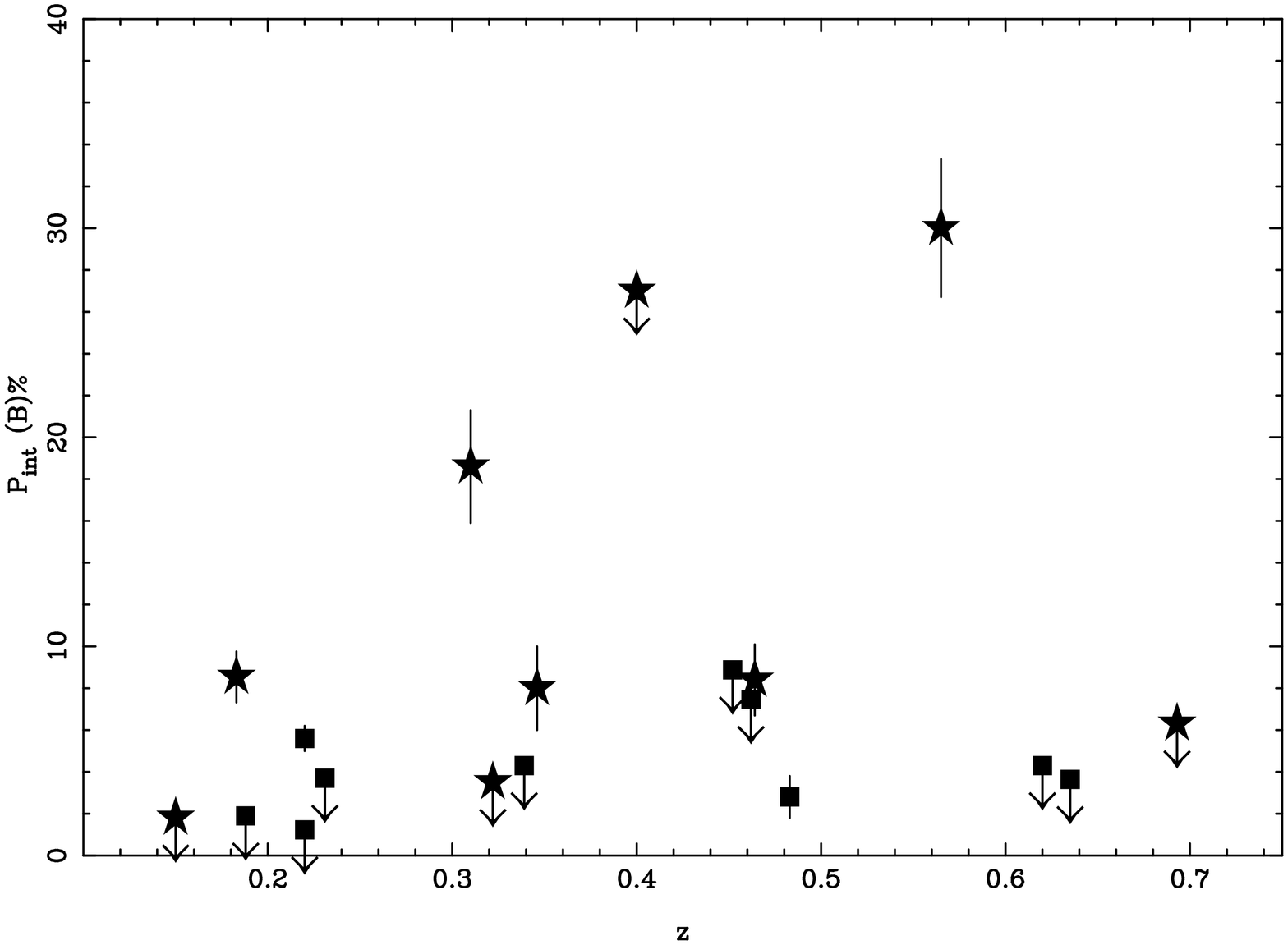,width=8cm,angle=270}
\end{center}
\caption[Intrinsic polarization vs. redshift]
{Plot of the intrinsic polarization as a function of redshift for the sample in Table 1. The BLRG are
 indicated by filled squares, and the NLRG  are indicated by filled stars.}
\label{figure:int-pol-z-plot}
\end{figure*}

By assuming that all the polarization is due entirely to the power-law component (see section 
4.4), we can use the fractional contribution of the power-law in the B-band filter, as 
determined from our modelling, to estimate the intrinsic polarization of the polarized 
component. In this way we automatically correct for the contributions of the old stellar 
populations and the nebular continuum, which are assumed to be unpolarized. A further 
assumption implicit in this procedure is that the fractional contribution of the polarized 
component is the same in the (different) apertures used for the spectroscopic and 
polarimetric observations, which is not necessarily the case. The results, which have
also been corrected for emission line contamination in the filter
band-pass, are shown in Table
4. From this it can be seen that the high levels of intrinsic polarization estimated 
for PKS2250-41 and PKS0117-15(3C38) are consistent with scattered light comprising a large fraction 
of the power-law component. However, in most of the other cases, the relatively low 
levels of intrinsic polarization ($P_{int}$) suggest that other, unpolarized, sources of UV light may 
contribute to the component modelled as a power-law

Figure 4  shows the intrinsic polarization plotted as a function of redshift. Although the 
upper envelope defined by the three most highly polarized sources shows evidence for a 
rise in the intrinsic polarization with redshift, taking the sample as a whole, and including 
the upper limits, this trend does not appear to be significant.

The proportion of highly polarized sources in our sample 
is somewhat less than found in other studies (Tadhunter et al. 1992, Cimatti et al. 1993,
Vernet et al. 2001). Possible reasons for this 
include the following.
\begin{itemize}
\item {\bf Observational biases of previous studies.} Because the polarization measurements are 
difficult to make, previous surveys are likely to have been biased towards the most 
spectacular and luminous radio galaxies in a given redshift range. Moreover, null 
results may not be always be reported in the literature.
\item{\bf Selection frequency of the current survey.} The selection frequency of the 2Jy sample 
(2.7GHz) is higher than for most other samples. However, it is not clear that this will necessarily 
bias our sample against high UV polarization. Most of our sources would in any case 
have been selected in low frequency surveys.
\item{\bf Redshift dependent effects.} For a given density of scatterers the strength of the 
scattered component will depend on the luminosity of the central illuminating source. 
The radio power is strongly correlated with the redshift in flux limited samples such as 
the 2Jy sample. Also, since optical emission line luminosity is strongly correlated with radio 
power (e.g. Rawlings \& Saunders 1991, Tadhunter et al. 1998), it is likely that the illuminating quasars hidden in the cores of the radio 
galaxies will become more luminous as the redshift increases. Therefore, we might 
expect the scattered quasar component to become more significant as the redshift 
increases and the sources become more powerful. Part of the difference between the 
results of our survey and those of previous surveys of higher redshift objects may be 
due to this effect.
\end{itemize}
The only way to unambiguously distinguish between these possibilities would be to make an 
optically unbiased survey of a large sample of radio galaxies selected from low frequency
radio surveys 
at a single frequency, and spanning a wide range in redshift. However, if the last possibility
is correct we would expect the most highly polarized sources in our sample to be those with the 
strongest signs of quasar activity, as deduced from the narrow emission line
luminosities. Figure 5 shows a plot of [OIII] emission line luminosity against radio power
for the sample in Table 1, with the significantly polarized objects
highlighted. As discussed in Tadhunter et al. (1998), [OIII] is more sensitive to
quasar ionizing luminosity than lower ionization emission lines such as
[OII](3727) and H$\beta$, and it is likely that
the large scatter in this plot is a consequence of a large
range in the quasar luminosity for a given radio power.
It is clear that many of the  highly polarized sources fall close to the upper envelope of 
the [OIII] luminosity in this plot, thus suggesting that the objects
with the most luminous quasars are also those with the
greatest proportional contribution of scattered AGN light in the UV.

\begin{figure} 
\begin{center}
\epsfig{file=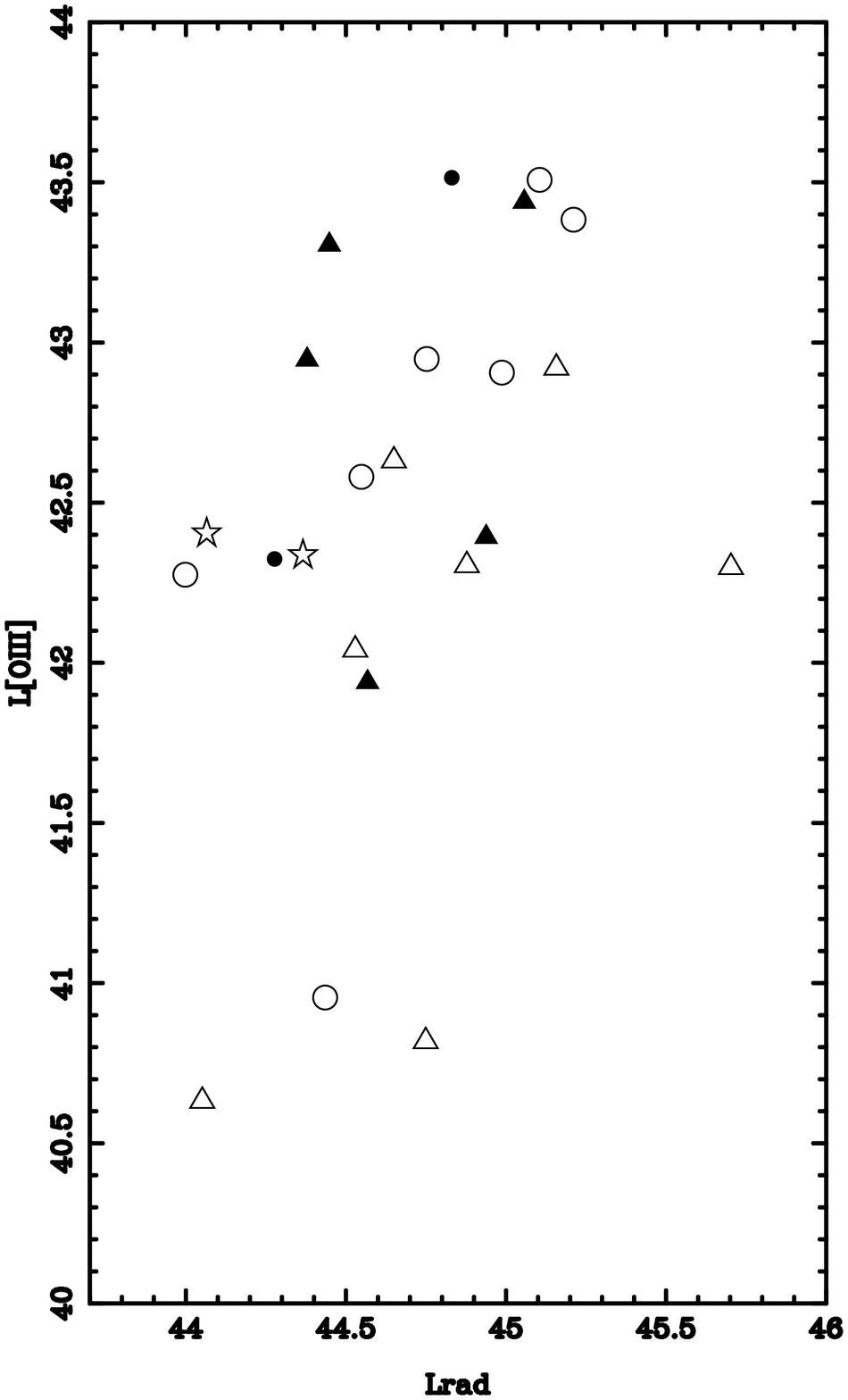,width=6.5cm,angle=0}
\end{center}
\caption[Emission line luminosity vs. total radio power.]
{[OIII] emission line luminosity plotted against total radio power for the complete sample in Table 1.
NLRG are indicated by triangles, while BLRG are idiciated by circles. Filled symbols indicate
objects for which a significant UV polarization has been detected. Stars indicate the two objects
in our sample (PKS1549-79 \& PKS2314+03) that are unusually bright at far-IR wavelengths.}  
\label{figure:d4000}
\end{figure}

\subsection{Direct AGN light}

We have identified 9 objects in our sample ($\sim$40\% of the total) which show 
broad permitted emission lines
in their spectra and should therefore be classified as BLRG\footnote{It is important
to distinguish between these BLRG --- in which the broad lines are readily
visible in the straight intensity spectra --- and objects in which the
broad lines are only visible in polarized spectra or following detailed
continuum modelling and subtraction (e.g. Cohen et al. 1999).}. With the exception
of PKS0035-02 (3C17), and possibly also PKS1547-79, none of the BLRG in our sample is highly
polarized in the UV, despite the presence of large UV excesses.
Therefore, it is likely that we are observing the nuclei directly in these sources,
and that the UV excess is due to direct AGN light,
with only a minor contribution, if any, from scattered AGN light. This is in line
with polarimetric observations of low redshift BLRG and quasars which show low levels
of polarization in most objects (Stockman, Angel \& Miley 1979, Antonucci 1984). 
Clearly, objects such as  3C234 (Tran, Cohen \& Goodrich 1995) and 3C109 
(Goodrich \& Cohen 1992), in 
which the broad lines
are readily detected in optical spectra, and which also show a high degree of linear
polarization, are the exception rather than the rule.

It is interesting to consider how these BLRG fit into the orientation-based unified 
schemes for powerful radio galaxies,
in which it is proposed that radio galaxies and quasars are the same thing viewed
from different directions, with the quasar nucleus blocked from direct view in the
radio galaxies by a central obscuring torus (Barthel 1989). There are two main 
possibilities:
\begin{itemize}
\item {\bf They are partially obscured quasars.} In this case, the quasar nuclei
would be observed at intermediate angles, such that they suffer mild extinction 
from the outer layers of the torus or the kpc-scale dust lane.
\item {\bf They are low luminosity quasars.} In this case, there would be a wide
range of quasar luminosity at a given radio power --- as
already suggested by the large scatter in the correlation between
[OIII] emission line luminosity and radio power (Tadhunter et al. 1998) --- and the BLRG would represent
the low luminosity end of the quasar luminosity function.
\end{itemize}
It would be difficult to distinguish between these possibilities based
on the current data alone, but the relatively red slopes of the power-laws fitted to the
BLRG in the two component model fits (see section 4.4) are more consistent with
the first of these two possibilities. The partial obscuration idea is also
consistent with the recent spectroscopic studies of quasars which suggest that a significant
proportion of quasars are significantly reddened (e.g. Baker 1997) at
optical wavelengths. Moreover, analyses of the radio properties of the 2Jy sample,
in particular the ratio of core to extended radio power, are consistent with the
idea that the BLRG are observed with the radio axis at an intermediate angle to
the line of sight (Morganti et al. 1997).


\subsection{Young stellar populations}

One of the most striking results from our survey is the emergence of a group of three 
radio galaxies (15\% of the total) in which the optical/UV continua are dominated by the light 
of young stellar populations. Interestingly, these three radio galaxies --- 3C459, 
PKS1549-79, PKS2135-20 --- all have relatively compact radio sources: 3C459 has an 
extended double radio source on a scale of 48 kpc (Morganti et al. 1999), but also has a 
steep spectrum radio core, which is itself double; PKS2135-20 is classified as a compact 
steep spectrum radio source; and, although PKS1549-79 has a flat radio spectrum and a one sided
jet structure extending to radial distances of 500pc, we have argued elsewhere on the basis of its 
emission line kinematics that this is an intrinsically compact radio source (Tadhunter et 
al. 2001). 

It is relatively easy to detect the young stellar populations in PKS2135-20, PKS1549-79 
and 3C459 because their optical continua are dominated by the young stars, and the ages 
of the starbursts (0.1 --- 2Gyr) are such that the spectral features of the 
young stellar population are relatively easy to detect. However, it is possible that 
significant young stellar populations are present in other radio galaxies in our sample, but 
are more difficult to detect because the stellar populations are younger and/or are less 
luminous relative to the old stellar populations. In particular, there are three sources in 
our sample with large UV excesses
--- PKS0023-26, PKS0409-75, PKS2211-17 ---  for which we have failed to identify the cause of the 
UV excess, because they show neither significant UV polarization nor broad permitted 
lines, and the nebular continuum makes a relatively minor contribution to their UV 
emission. There are a further three objects --- PKS0039-44, PKS1934-63, PKS1306-09 --- which are 
significantly polarized in the UV, but the level of intrinsic polarization, if attributed 
solely to the power-law component, is less than expected for pure scattered light. Finally,
we note that the residuals of the two component fits to the continuum of the weak broad line
object  PKS1932-46 --- with a significant excess in the 3800 -- 4300\AA\, region
and a blue continuum slope at longer wavelengths --- are consistent with with the 
presence of a young stellar population. Although young
stellar populations may make a significant contribution in all of these objects, the relatively red
slopes of the fitted power-laws for most of them (Tables 7 \& 8) suggest that
the starburst component must be significantly reddened.

\begin{figure} 
\begin{center}
\epsfig{file=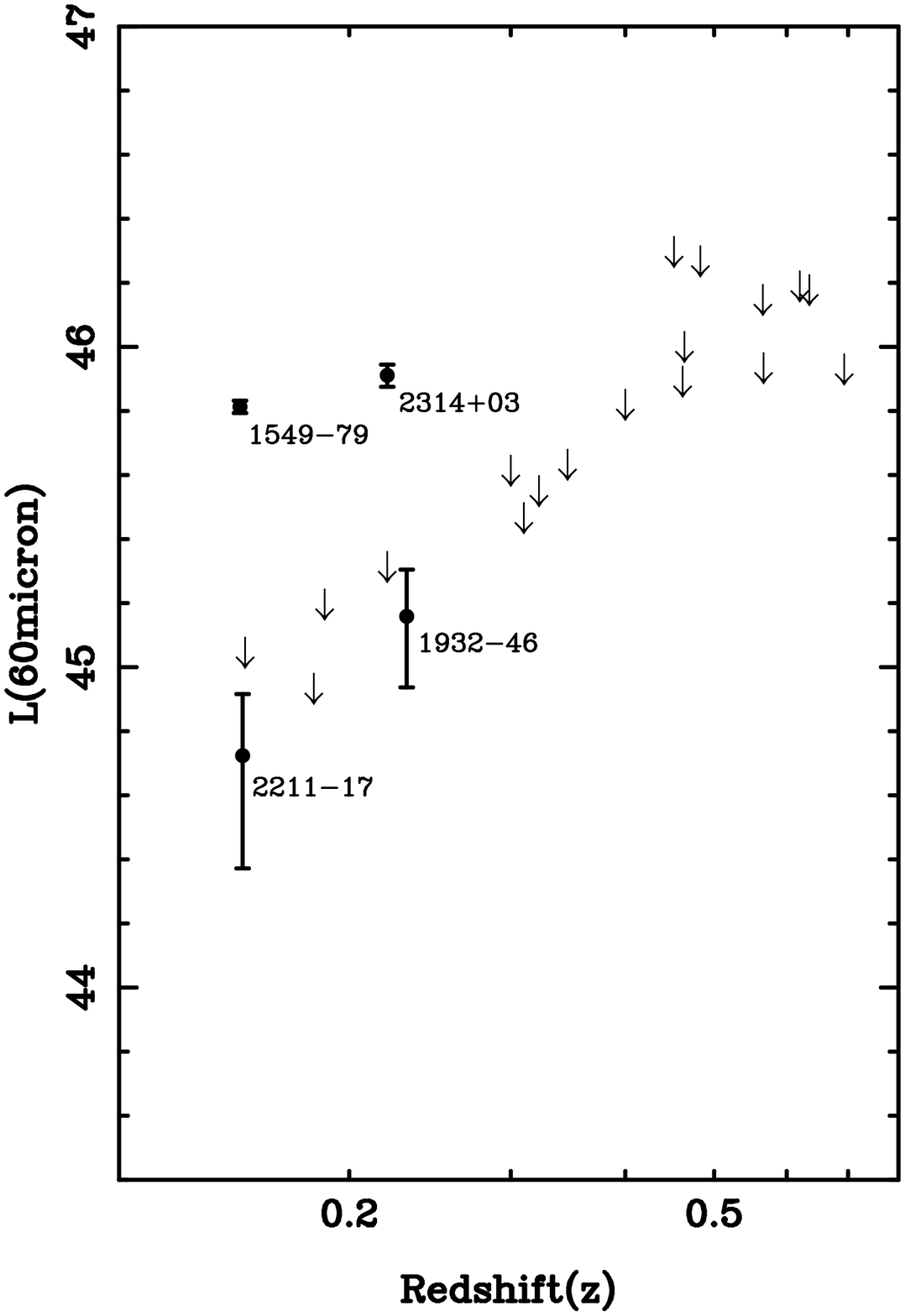,width=7.5cm,angle=0}
\end{center}
\caption[Far-IR luminosity plotted against redshift]
{A plot of far-infrared (60$\mu$m) luminosity against redshift.}  
\label{figure:farir}
\end{figure}

Taking these six less certain cases along with three objects in which the young stellar 
populations dominate, we derive a conservative upper limit of 50\% on the proportion of 
objects in our sample which have their optical/UV continua dominated by 
the light of young stellar populations. This is entirely consistent with the results
from two recent surveys of lower redshift 3C radio sources in which $\sim$30 -- 40\% of the objects
show evidence for recent star formation activity (Aretxaga et al. 2001, Wills et al. 2001).

\subsection{The far-IR/starburst link in powerful radio galaxies}
 
The far-IR/sub-mm 
continuum excesses
detected in some radio galaxies are often attributed to dust heated by starbursts
(e.g. Rowan-Robinson 1995, Archibald et al. 2001). However,  this
interpretation is controversial, given the likely presence of powerful quasar nuclei hidden
in the cores of the galaxies which could also heat the cooler dust component
(Sanders et al. 1989). Indeed,
the issues surrounding the nature of the far-IR continuum are similar to those
surrounding the UV excess considered in this paper: is the far-IR/sub-mm  excess primarily
a direct consequence of the quasar activity, or is it a consequence of starbursts related
to the evolutionary events that triggered the activity?

In order to investigate the links between the far-IR excess and optical/UV starburst
activity, we have extracted 60$\mu$m fluxes and 3$\sigma$ upper limits from IRAS `all-sky' 
survey data. To do this we used the SCANPI algorithm available at IPAC to co-add individual
IRAS scans. The IRAS co-added scans were then visually inspected in order to assess
the significance of individual detections. The detection of an object was only considered significant if
the positional offset in the detection was less than 0.5 arcminutes from the nominal
position of the object, and the object was detected in more than one IRAS band.
In cases without clear detections we quote 3$\sigma$ upper limits, where $\sigma$ was
taken as the RMS deviation from the baseline-corrected scans at 60$\mu$m.  The resulting 
fluxes and upper limits are presented in Table 10. 
We have also determined the 60$\mu$m luminosities ($\nu P_{\nu}$), by assuming a
far-IR spectral slope of -1 (see Golombek et al. 1988). The luminosities are
presented in the last column of Table 10, and plotted against redshift in Figure 6.
It is notable that only 4 out of 22 of the sources
in our sample were detected by IRAS at 60$\mu$m with any certainty. 

\begin{table} 
\begin{center}
\begin{tabular}{lll} \hline \hline
{\bf Object} & {\bf $F_{60\mu m}$ (Jy)} & {\bf $\nu P_{60\mu m}$ (erg s$^{-1}$)}\\ \hline
0023---26 & $< 0.117$ & $< 3.6\times 10^{45}$  \\
0035--02 & $< 0.159$ &$< 2.0\times 10^{45}$ \\
0038+09  & $< 0.171$ &$< 1.5\times10^{45}$   \\
0039--44 & $< 0.120$   & $< 4.3\times 10^{45}$ \\
0105--16 & $< 0.132$ & $< 6.5\times 10^{45}$ \\
0117--15 & $< 0.123$ & $< 1.4\times 10^{46}$\\ 
0235--19 & $< 0.108$  & $< 1.6\times 10^{46}$\\
0252--71 & $< 0.075$  & $< 8.6\times 10^{45}$ \\ 
0347+05  & $< 0.159$  & $< 4.0\times 10^{45}$ \\
0409--75 & $< 0.045$ & $< 8.6\times 10^{45}$\\
1306--09 & $< 0.141$ & $< 9.9\times 10^{45}$\\
1547--79 & $< 0.237$ & $< 1.8\times 10^{46}$\\
1549--79 & $1.120 \pm 0.05$  & $(6.5 \pm 0.3)\times10^{45}$\\
1602+01  & $<0.111$ & $< 7.9\times 10^{45}$\\ 
1648+05 & $< 0.186$  & $< 1.1\times 10^{45}$ \\
1932--46 & $0.100  \pm 0.04$ & $(1.4 \pm 0.6)\times10^{45}$\\
1934--63 & $< 0.099$  & $< 8.6\times 10^{44}$ \\
1938--15 & $< 0.297$ & $< 2.0\times 10^{46}$ \\
2135--20 & $< 0.099$ & $< 1.5\times 10^{46}$ \\
2211--17 & $0.090 \pm 0.05$  &$(5.4 \pm 1.9)\times10^{44}$ \\
2250--41 & $< 0.105$ & $< 2.9\times 10^{45}$ \\
2314+03  & $0.630 \pm 0.05$ & $(8.1 \pm 0.7)\times10^{45}$\\
\end{tabular} 
\end{center}
\caption{Far-infared (60$_{\mu m}$) fluxes and luminosities for the
sample in Table 1, as derived from co-added
IRAS scans. The $\nu P_{60\mu m}$ luminosities were derived by assuming
a far-IR continuum spectral index of $\beta = -1.0$ ($F_{\nu} \propto \nu^{\beta}$) over the
wavelength range 25 -- 100$\mu m$.}
\end{table}

The most striking feature of the far-IR luminosities in Table 10 and Figure 6 is that
two of the three objects in our sample dominated by optical/UV starbursts --
PKS1549-79 \& PKS2314+03 (3C459) -- also stand out as being unusually luminous at far-IR
wavelengths. Indeed,  PKS1549-79 \& PKS2314+03 (3C459) are the most far-IR 
luminous  of the 12 objects in our sample at $z < 0.40$; they
are approximately a factor 5 -- 10$\times$ brighter at far-IR wavelengths than the  other
2Jy radio galaxies
at similar redshifts. The third object dominated by an
optical/UV starburst -- PKS2135-20 --
is at much higher redshifts and its 60$\mu$m upper limit is consistent with it having a
far-IR luminosity similar to those of PKS1549-79 \& PKS2314+03. Note the [OIII] emission
line luminosities of PKS1549-79 \& PKS2314+03  are typical of radio galaxies at similar
redshifts; they are not in any way extreme (see Figure 5). Therefore
it is unlikely that the far-IR excesses result from heating of the
cool dust by unusually luminous quasars, or that the covering factor of gas and dust 
is unusually large in these objects.   

PKS1932-46 \& PKS2211-17 were also marginally detected by IRAS. Again, these are both sources
in which significant young stellar populations are suspected, based on their optical/UV
spectra (see section 5.3).

Overall, despite the low rate of detection of the radio galaxies in our sample at
far-IR wavelengths, there appears to be a  link between the detection of
optical/UV starburst activity and the level of the far-IR excess. Recently,
we obtained a  similar result for a sample of lower redshift 3C sources
(Wills et al. 2001): the object with the largest
contribution from an optical/UV starburst in that sample (3C433) also has
the largest far-IR luminosity. All of
these results support the
idea that the cooler dust components detected in radio galaxies are heated by starbursts
rather than hidden quasar nuclei. The putative link between far-IR excess and optical/UV starburst
activity for powerful radio galaxies will be discussed in more detail in a future
paper (Tadhunter et al. 2002, in preparation). 

\subsection{Implications for studies of quasar host galaxies}

If the unified schemes for powerful radio galaxies are correct, then radio galaxies {\it are}
quasar host galaxies in which  the quasar nuclei are hidden from our direct view. Therefore, we expect
to find the same components contributing to the UV excess in the quasar hosts as we do
in the powerful radio galaxies. Indeed, if the scattering dust grains are large relative
to the wavelength of the optical/UV light, the scattered AGN component may be more significant
in the quasar hosts, because of the forward scattering properties of the dust grains. 
On the basis the results reported above, it would be unwise to base any investigation of quasar
host galaxy properties on continuum colours alone. Clearly, detailed spectroscopic and
polarimetric observations are as important for disentangling the various components contributing
to the optical/UV continuum in quasar hosts, as they are in powerful radio galaxies.

\section{Conclusions and further work}

The results of our survey confirm the multi-component nature of the UV continuum in
powerful radio galaxies. We can quantify the contributions of the various
components that contribute to the UV excess in the near-nuclear regions as follows:
\begin{itemize}
\item {\bf Nebular continuum.} This is present in the spectra of all the
objects and contributes 
5 -- 40\% of the UV continuum below the Balmer break. 
\item {\bf Direct AGN light.} Based on the detection of broad permitted lines
in the intensity spectra, direct AGN light makes a significant contibution to
the UV excess in $\sim$30 -- 40\% of the objects in our sample.
\item {\bf Scattered AGN light.} Polarimetric observations provide evidence that
scattered AGN light makes a significant contribution to the UV continuum in 
37\% of the 19 objects in our sample with polarization measurements, but in most
cases the scattered component does not appear to dominate the UV excess.
\item {\bf Starburst component.} Young stellar populations clearly dominate
the optical/UV continua in three of the objects in our sample
($\sim$15\%) and may make a large contribution to the optical/UV continuum 
in up to 50\% of all
the sample objects.
\end{itemize}

Of these components, the starburst component clearly warrants further investigation. If, as
seems likely, the starbursts were triggered by the same merger events that triggered
the radio jet and quasar activity, then detailed studies of the starbursts can provide
potentially unique information about the genesis of powerful radio sources, particularly 
the timescales, the nature of the mergers and the order-of-events.  

\subsection*{Acknowledgments} Based on observations collected at
the European Southern Observatory, La Silla, Chile, and at 
the Roque de los Muchachos Observatory, La Palma.
RD,MVM,TGR \& KAW acknowledge support from PPARC while this
work was being carried out.
{}
\end{document}